\documentclass[sigconf]{acmart}
\usepackage{float}
\usepackage{multirow}
\usepackage{CJKutf8}
\usepackage{booktabs}
\usepackage{graphicx}
\usepackage{amsmath}
\usepackage{bm}
\usepackage{algorithmic}
\usepackage[ruled,linesnumbered]{algorithm2e}
\usepackage{amsfonts}
\usepackage{caption}
\usepackage{etoolbox}
\usepackage{makecell}
\usepackage{tabularx}
\usepackage{enumitem}
\usepackage{float}
\usepackage{subfig}
\usepackage{placeins}
\usepackage{pifont}
\usepackage[normalem]{ulem}
\newcommand{\cmark}{\ding{51}}
\newcommand{\pmark}{\ensuremath{\triangle}}
\newcommand{\xmark}{\ding{55}}
\DeclareRobustCommand{\unirank}{\texorpdfstring{\textsf{UniRank}}{UniRank}}
\providecommand{\symbfit}[1]{\bm{#1}}
\AtBeginEnvironment{figure}{%
  \captionsetup{justification=raggedright, font=small, skip=2pt}%
}
\AtBeginEnvironment{figure*}{%
  \captionsetup{justification=raggedright, font=small, skip=2pt}%
}
\AtBeginEnvironment{table}{%
  \setlength{\abovecaptionskip}{0.1cm}%
  \setlength{\belowcaptionskip}{0.1cm}%
  \captionsetup{font=small}%
}
\AtBeginEnvironment{table*}{%
  \setlength{\abovecaptionskip}{0.1cm}%
  \setlength{\belowcaptionskip}{0.1cm}%
  \captionsetup{font=small}%
}
\renewcommand\footnotetextcopyrightpermission[1]{} % remove copyright notice
\AtBeginDocument{%
  }

\setcopyright{acmlicensed}
\copyrightyear{2018}
\acmYear{2018}
\acmDOI{XXXXXXX.XXXXXXX}
\acmConference[Conference acronym 'XX]{Make sure to enter the correct
  conference title from your rights confirmation email}{June 03--05,
  2018}{Woodstock, NY}
\acmISBN{978-1-4503-XXXX-X/2018/06}

\begin{document}

\title[\unirank: Benchmarking Ranking Models for Unified Sequential Modeling and Feature Interaction]{%
  \raisebox{-0.28\height}{\includegraphics[height=1.35em]{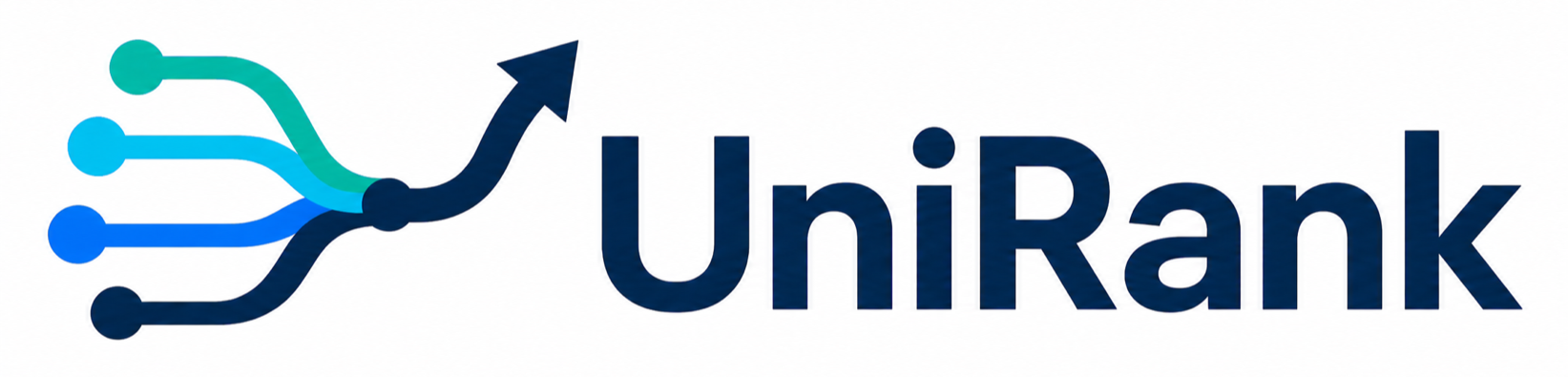}}: Benchmarking Ranking Models for Unified Sequential Modeling and Feature Interaction%
}

\author{Honghao Li}
\email{salmon1802li@gmail.com}
\orcid{0009-0000-6818-7834}
\affiliation{%
  \institution{Anhui University}
  \city{Hefei}
  \state{Anhui Province}
  \country{China}
}

\author{Xianquan Wang}
\email{wxq2001@gmail.com}
\orcid{0009-0003-9744-220X}
\affiliation{%
  \institution{Independent Researcher}
  \city{Hefei}
  \state{Anhui Province}
  \country{China}
}

\author{Zibin Zhang}
\email{bingoozhang@tencent.com}
\affiliation{
  \institution{Tencent Inc.}
  \city{Guangzhou}
  \state{Guangdong Province}
  \country{China}
}

\author{Yi Zhang}
\email{zhangyi.ahu@gmail.com}
\affiliation{%
  \institution{Anhui University}
  \city{Hefei}
  \state{Anhui Province}
  \country{China}
}

\author{Kangyi Lin}
\email{konyel@163.com}
\affiliation{
  \institution{Tencent Inc.}
  \city{Guangzhou}
  \state{Guangdong Province}
  \country{China}
}

\author{Yiwen Zhang}
\authornote{Corresponding author}
\email{zhangyiwen@ahu.edu.cn}
\affiliation{
  \institution{Anhui University}
  \city{Hefei}
  \state{Anhui Province}
  \country{China}
}

\renewcommand{\shortauthors}{Anonymous et al.}

\begin{abstract}
Ranking is a core stage in online advertising and recommender systems. Modern ranking models increasingly unify sequential modeling and feature interaction, yet many advances rely on proprietary data, closed implementations, and large-scale industrial infrastructure. This setting limits reproducible comparison and hinders academic study of scaling laws, long-sequence modeling, and multi-task ranking. To address these limitations, this paper proposes \textbf{\unirank}, an open benchmark for ranking models that unify sequential modeling and feature interaction. \unirank{} uses chronological pointwise autoregressive supervision, standardizes evaluation across feedback tasks, and provides a PyTorch toolkit with Distributed Data Parallel training, operator optimization, mixed-precision training, attention optimization, and other efficiency techniques that reduce hardware requirements. We benchmark 15 representative unified ranking models on five public datasets from short-video, advertising, and e-commerce platforms, with the largest containing over 700 million instances and the longest behavior sequence exceeding $10^5$ interactions. \unirank{} provides a reproducible foundation for comparing unified ranking models, studying scaling laws under limited compute, and narrowing the gap between academic and industrial ranking research. We believe \unirank{} benefits researchers, practitioners, and beginners through reproducible experiments, production-oriented evaluation, and accessible implementations. Code and data are available at \url{https://github.com/salmon1802/UniRank}.
\end{abstract}


\maketitle

\section{Introduction}
Recommender systems drive user engagement and business value in e-commerce~\cite{DIN,DIEN}, short-video streaming~\cite{kuairand}, and online advertising~\cite{dcn,widedeep}. Under large item corpora and strict latency constraints, industrial recommender systems typically adopt a multi-stage pipeline, including \textit{retrieval}~\cite{DNN,mind}, \textit{pre-ranking}~\cite{cold,gu2022ranking}, \textit{ranking}~\cite{deepfm,xdeepfm}, and \textit{re-ranking}~\cite{ai2018learning,prm}.

% \begin{figure}[h]
%   \centering
%   \includegraphics[width=0.75\linewidth]{figures/workflow.pdf}
%   \caption{A workflow of modern recommender systems.}
%   \label{fig:workflow}
% \end{figure}

Within this pipeline, the ranking model is the central decision component~\cite{deepfm}. Given heterogeneous multi-field features such as user profiles, item attributes, device types, request times, and historical behaviors, the ranking model estimates the probabilities of task-specific feedback events, including clicks, conversions, and other user responses~\cite{openbenchmark,FuxiCTR,xdeepfm}. Consequently, ranking models directly affect user experience and platform revenue, and even small improvements in offline metrics such as the area under the ROC curve (AUC) can correspond to meaningful business gains~\cite{widedeep}. The design and optimization of ranking models therefore remain important in both academia and industry~\cite{DIN,DIEN}.

Recent advances push ranking models toward more expressive feature representations and user intent signals. This progression yields three broad methodological families: (1) \textit{The Shallow Learning Method} uses Logistic Regression~\cite{LR} and factorization-based models~\cite{FM,FFM,FMFM}. These models are efficient but have limited capacity and rely on manual feature engineering. (2) \textit{The Deep Feature Interaction Method} embeds sparse categorical features and replaces manual crosses with neural interaction modules. Representative examples use MLPs, compressed interaction networks, cross networks, or quadratic neurons to learn nonlinear high-order feature combinations, including DeepFM~\cite{deepfm}, xDeepFM~\cite{xdeepfm}, DCNv2~\cite{dcn}, and QNN~\cite{QNN}. (3) \textit{The Sequence Modeling Method} captures temporal changes in user interest. Models such as DIN~\cite{DIN} and DIEN~\cite{DIEN} use target-aware attention or recurrent networks to aggregate histories for the target item. To improve performance, deployed ranking models~\cite{DLRM,DHEN} may integrate multiple DCNv2, MLP, and DIN modules. However, these non-unified designs introduce fine-grained operators and execution boundaries that limit parallel scaling~\cite{rankmixer}.

Inspired by large language models~\cite{onerec_report,rankmixer}, current research increasingly adopts token-based unified architectures that combine sequential modeling with feature interaction. These architectures improve Model FLOPs Utilization (MFU) and scale more effectively with sequence length and model capacity. They follow either \textit{stacked} designs~\cite{hiformer,rankmixer,zenith,tokenmixer,unimixer,hemix,ssr}, which perform sequential modeling before feature interaction, or \textit{layer-wise} designs~\cite{onetrans,hyformer,mixformer_rec,infnet,est,tokenformer,longer,ultrahstu}, which integrate both operations within each layer. Together, these architectures provide a unified foundation for studying how sequential modeling and feature interaction scale under a consistent protocol.

Despite this progress, scaling ranking research to industrial conditions remains challenging. Many strong ranking models, large-scale datasets, and training systems rely on proprietary resources, leaving academic researchers with limited access to the data and infrastructure needed to study scaling laws, long-sequence behavior, and task trade-offs under realistic conditions. These barriers constrain academic exploration and systematic comparison.

Existing studies and toolkits provide integrated platforms for recommendation evaluation, including FuxiCTR~\cite{openbenchmark,FuxiCTR}, Scenario-Wise Rec~\cite{scenariowiserec2025}, DeepCTR~\cite{shen2017deepctr}, RecBole~\cite{zhao2021recbole}, and MMLRec~\cite{yuan2024mmlrec}. These frameworks greatly facilitate standardized research across complementary settings: FuxiCTR and DeepCTR focus on single-task CTR prediction, RecBole covers general and sequential recommendation, and Scenario-Wise Rec and MMLRec emphasize multi-scenario or multi-task learning. However, jointly extending this standardization to industrial-scale long-sequence modeling and unified architectures remains challenging. In particular, no common protocol yet exists for chronological data partitions, behavior sequence construction, capacity-controlled comparison, and evaluation across feedback tasks. Further evidence is needed to understand how sequential modeling and feature interaction scale across feedback signals.

To address these issues, we propose the Unified Ranking Model Benchmark (\textbf{\unirank}) for ranking models with unified sequential modeling and feature interaction. Specifically, \unirank{} evaluates 15 representative models on five large-scale public datasets from short-video, advertising, and e-commerce platforms. It also provides handbook-style controlled experiments on tokenization, activation functions, architectural enhancements, optimizers, sequence length, and model scale. These experiments standardize architectural studies and reduce the cost of rebuilding datasets, implementations, and ablation studies.

\enlargethispage{2\baselineskip}

This work serves several groups of readers:
\begin{itemize}[leftmargin=*,nosep]
\item \textbf{Researchers}: \unirank{} provides accessible open-source code and processed datasets for reproducible model development, controlled experimentation, and~\mbox{direct architectural comparison} across standardized settings.
\item \textbf{Practitioners}: \unirank{} provides acceleration components, supports multi-feedback tasks, and offers production-oriented recipes for efficient training and practical validation.
\item \textbf{Beginners}: The repository serves as an educational resource with accessible PyTorch implementations and configurations that help readers understand~\mbox{modern unified ranking architectures} through hands-on examples.
\end{itemize}

In summary, our work makes the following main contributions:
\begin{itemize}[leftmargin=*,nosep]
\item We propose \textbf{\unirank}, the first open benchmark dedicated to ranking architectures that unify sequential modeling and feature interaction, enabling systematic comparison under standardized data, optimization, and efficiency settings.
\item We integrate Distributed Data Parallel (DDP), operator, mixed-precision, and attention optimizations to reduce hardware costs.
\item We release source code, training configurations, processed datasets, and complete results for 15 models on five public datasets, spanning up to 700 million instances and behavior sequences exceeding $10^5$ events, supporting reproducible evaluation and future benchmark extensions.
\end{itemize}

\section{Ranking Models}
\subsection{Input Features and Modeling Paradigms}
Ranking models estimate target-item feedback probabilities. \unirank{} represents each impression with non-sequential features $\mathcal{F}$, behavior sequences $\mathcal{S}$, and target features $\mathcal{V}$:
\begin{equation}
\mathcal{F}=\{f_1,\ldots,f_m\}, \quad
\mathcal{S}=(s_1,\ldots,s_L), \quad
\mathcal{V}=\{v_1,\ldots,v_n\},
\end{equation}
where $\mathcal{F}$ contains non-sequential user and context fields, such as \textit{user ID, age group, membership level, device type, request time, and traffic source}. $\mathcal{S}$ is the chronological behavior sequence and includes \textit{historical item-side features and action types, such as category, creator or seller, brand, and other available metadata}. $\mathcal{V}$ contains the current target-item features. During training, the latest behavior in each instance defines the target and feedback label, while $\mathcal{S}$ contains only preceding behaviors. After embedding and tokenization, $\symbfit{H}_{\mathcal{S}}$ and $\symbfit{H}_{\mathcal{F},\mathcal{V}}$ denote the sequence and joint non-sequential--target representation matrices. The following three paradigms progressively incorporate sequence representations into prediction.

\subsubsection{\textbf{Traditional Non-Sequential Modeling Paradigm}}
This paradigm models interactions between non-sequential heterogeneous features and target features:
\begin{equation}
\hat{y}=\phi_{\mathrm{ns}}\!\left(\symbfit{H}_{\mathcal{F},\mathcal{V}}\right).
\end{equation}
This formulation covers shallow and deep subnetworks that learn feature crosses from $\symbfit{H}_{\mathcal{F},\mathcal{V}}$ without modeling $\symbfit{H}_{\mathcal{S}}$. DeepFM~\cite{deepfm} and DCNv2~\cite{dcn} are representative examples. Because they omit temporal dependencies, \unirank{} treats them as conventional baselines outside the unified benchmark.

\subsubsection{\textbf{Stacked Unified Interaction}}
This paradigm first aggregates $\symbfit{H}_{\mathcal{S}}$ and then combines it with $\symbfit{H}_{\mathcal{F},\mathcal{V}}$:
\begin{equation}
\symbfit{H}_{g}=g_{\mathrm{agg}}\!\left(\symbfit{H}_{\mathcal{S}},\symbfit{H}_{\mathcal{F},\mathcal{V}}\right),\quad
\hat{y}=\phi_{\mathrm{stack}}\!\left(\symbfit{H}_{\mathcal{F},\mathcal{V}},\symbfit{H}_{g}\right),
\end{equation}
where $\symbfit{H}_{g}$ is the aggregated sequence representation matrix and $\phi_{\mathrm{stack}}$ is the downstream interaction and prediction network. DIN~\cite{DIN} provides an early target-aware pooling example but lacks a token-based unified architecture and is excluded. \unirank{} evaluates token-based models such as HiFormer~\cite{hiformer}, RankMixer~\cite{rankmixer}, and UniMixer~\cite{unimixer}.

\subsubsection{\textbf{Layer-wise Unified Interaction}}
This paradigm exchanges information between sequential and non-sequential representations within every interaction layer:
\begin{equation}
(\symbfit{H}_{\mathcal{S}}^{(l+1)},\symbfit{H}_{\mathcal{F},\mathcal{V}}^{(l+1)})=
g^{(l)}(\symbfit{H}_{\mathcal{S}}^{(l)},\symbfit{H}_{\mathcal{F},\mathcal{V}}^{(l)}),\quad
\hat{y}=\phi_{\mathrm{out}}(\symbfit{H}^{(L)}),
\end{equation}
where $l\in\{0,\ldots,L-1\}$ indexes the interaction layers, $\symbfit{H}_{\mathcal{S}}^{(l)}$ is the sequence representation matrix at layer $l$, and~$\symbfit{H}_{\mathcal{F},\mathcal{V}}^{(l)}$ is the joint representation matrix of non-sequential and target tokens. The function $g^{(l)}$ performs unified interaction within layer $l$, $\symbfit{H}^{(L)}$ is the final representation matrix, and $\phi_{\mathrm{out}}$ maps it to $\hat{y}$. Unlike stacked architectures, this formulation preserves behavior, field, and target representations across layers and supports iterative information exchange. OneTrans~\cite{onetrans}, TokenFormer~\cite{tokenformer}, and UltraHSTU~\cite{ultrahstu} instantiate this paradigm with distinct interaction operators.

\subsection{The Evolution of Training Paradigms}
\subsubsection{\textbf{Traditional Single-Task New-Impression-Only Paradigm}}
\begin{figure}[t]
  \centering
  \includegraphics[width=0.75\linewidth]{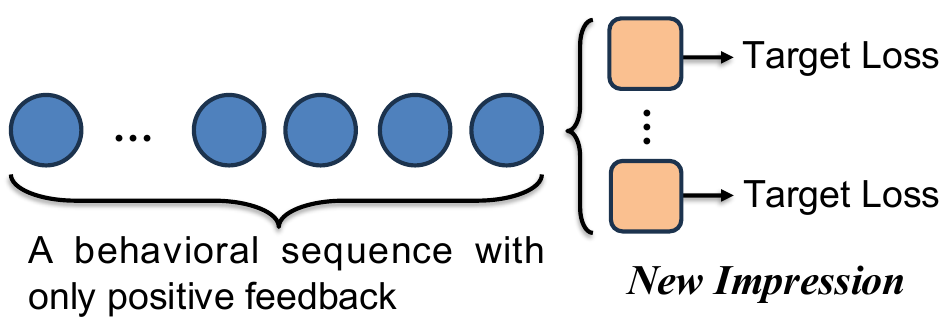}
  \caption{Traditional single-task New-Impression-Only paradigm.}
  \label{fig:train-past}
  \vspace{-0.5em}
\end{figure}

Traditional ranking pipelines retain only positive behaviors in the user history and use this history to predict a new impression. Figure~\ref{fig:train-past} treats the behavioral sequence as fixed auxiliary context: each new impression forms an independent sample and contributes one loss for a single feedback task, typically click prediction. The objective is
\begin{equation}
\mathcal{L}_{\mathrm{NIO}}=\sum_{i=1}^{N}\mathrm{BCE}\!\left(y_i,
f_{\theta}(\mathcal{S}_i^{+},\symbfit{x}_i)\right),
\label{eq:nio}
\end{equation}
where $N$ is the number of new-impression samples, $\mathcal{S}_i^{+}$ is the fixed positive-feedback sequence shared by multiple samples including $i$, $\symbfit{x}_i$ is the new-target feature vector, $y_i$ is its binary label, and $f_{\theta}$ is the ranking model with parameters $\theta$. For each sample $i$, Equation~\ref{eq:nio} applies the loss only to the new target $\symbfit{x}_i$. Events in $\mathcal{S}_i^{+}$ remain fixed auxiliary context and never become target positions, so they have neither labels nor loss terms and receive no direct per-position supervision. Consequently, gradients arise only from new targets, leaving sequences sparsely supervised.

\begin{figure}[t]
  \centering
  \includegraphics[width=0.75\linewidth]{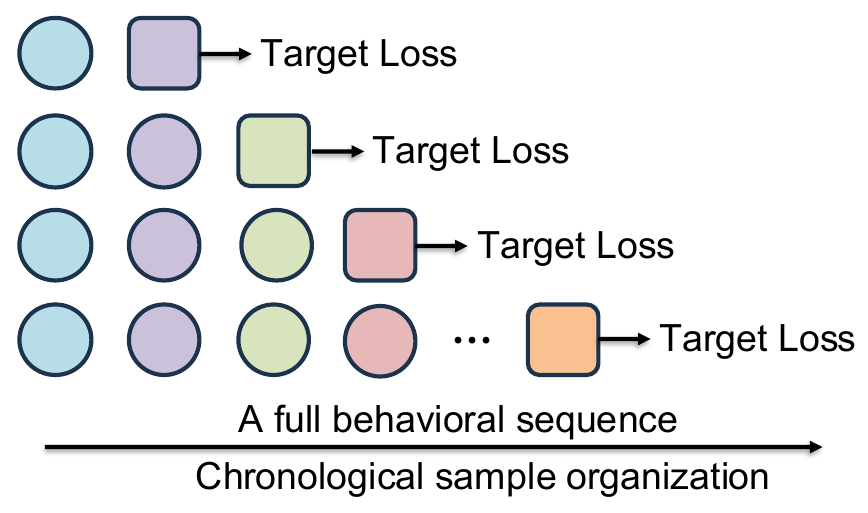}
  \caption{Pointwise autoregressive paradigm.}
  \label{fig:train-ours}
  \vspace{-0.5em}
\end{figure}

\begin{table*}[t]
\Huge
\renewcommand\arraystretch{1.2}
\caption{Reproducibility requirement statistics.}
\resizebox{\linewidth}{!}{
\begin{tabular}{c|cccccccccccccccc}
\hline
\textbf{Reproducibility requirements}        & \textbf{HiFormer} & \textbf{RankMixer} & \textbf{Zenith} & \textbf{TokenMixer} & \textbf{UniMixer} & \textbf{HeMix} & \textbf{SSR} & \textbf{OneTrans} & \textbf{LONGER} & \textbf{HyFormer} & \textbf{MixFormer} & \textbf{INFNet} & \textbf{EST} & \textbf{TokenFormer} & \textbf{UltraHSTU} & \textbf{Ours} \\ \hline
\textbf{Unified data pipeline}               & \pmark & \pmark & \pmark & \pmark & \pmark & \pmark & \pmark & \pmark & \pmark & \pmark & \pmark & \pmark & \pmark & \pmark & \pmark & \cmark \\
\textbf{Unified task and metric definitions} & \cmark & \cmark & \cmark & \cmark & \cmark & \cmark & \cmark & \cmark & \cmark & \cmark & \cmark & \cmark & \cmark & \cmark & \pmark & \cmark \\
\textbf{Unified model source code}           & \xmark & \xmark & \xmark & \xmark & \xmark & \xmark & \cmark & \xmark & \xmark & \xmark & \xmark & \xmark & \xmark & \xmark & \xmark & \cmark \\
\textbf{Unified model hyperparameters}       & \cmark & \pmark & \pmark & \pmark & \cmark & \pmark & \pmark & \cmark & \cmark & \cmark & \pmark & \cmark & \cmark & \pmark & \cmark & \cmark \\
\textbf{Unified training pipeline}           & \pmark & \pmark & \pmark & \pmark & \pmark & \pmark & \pmark & \cmark & \pmark & \pmark & \pmark & \pmark & \pmark & \pmark & \pmark & \cmark \\
\textbf{Unified efficiency optimization}     & \pmark & \pmark & \pmark & \pmark & \pmark & \pmark & \pmark & \cmark & \cmark & \pmark & \pmark & \pmark & \pmark & \pmark & \cmark & \cmark \\ \hline
\end{tabular}}
\label{reproducibility}
\vspace{-0.5em}
\end{table*}

\subsubsection{\textbf{Chronological Pointwise Autoregressive Paradigm}}
Figure~\ref{fig:train-ours} shows that \unirank{} constructs a chronological full-feedback sequence rather than a positive-only history, retaining positive actions and implicit negative outcomes such as exposed items without clicks~\cite{beyond_positive}. Each eligible position $t$ becomes an independent pointwise sample. The model predicts the current target's binary labels from the strictly preceding sequence $\mathcal{S}_{<t}$ of items, actions, and timestamps. Accordingly, a sample may activate multiple labels, and each model uses one prediction head per task. Its target loss sums binary cross-entropy terms:
\begin{equation}
\mathcal{L}_{\mathrm{AR}}=\sum_{t\in\mathcal{T}}\sum_{k=1}^{K}
\mathrm{BCE}\!\left(y_{t,k},f_{\theta}(\mathcal{S}_{<t},\symbfit{x}_t)\right),
\label{eq:ar}
\end{equation}
where $\mathcal{T}$ is the set of eligible target positions, $K$ is the number of feedback tasks, $\mathcal{S}_{<t}$ is the full-feedback event sequence before position $t$, $\symbfit{x}_t$ is the current-target feature vector, and $y_{t,k}$ is its label for task $k$. Equation~\ref{eq:ar} provides up to $|\mathcal{T}|K$ supervised loss terms and direct gradients from successive positions rather than only the newest target. Because the chronological sequence retains every eligible event type, negative outcomes become both contextual representations and supervised targets. The resulting dense gradients improve the use of long interaction sequences and scale with chronological data and feedback tasks. This paradigm therefore better supports scaling in sequence length, data volume, and model capacity, while action-aware sequences capture dependencies across feedback tasks.

\subsection{Representative Models}

\subsubsection{\textbf{Stacked Unified Interaction}}
This paradigm arranges sequential modeling and feature interaction as consecutive modules. The sequence module first encodes the behavioral history, and the feature interaction module then combines this representation with user, target-item, and context features.
\begin{itemize}[leftmargin=*]
\item \textbf{HiFormer}~\cite{hiformer} pools the behavior history with target attention, tokenizes heterogeneous fields and task queries, and applies hierarchical HiFormer blocks to the stacked tokens.
\item \textbf{RankMixer}~\cite{rankmixer} pools the history with target attention, tokenizes concatenated ranking fields into chunks, and alternates parameter-free Token Mixing with per-token MLPs.
\item \textbf{Zenith}~\cite{zenith} separately embeds user and item identifiers, pools historical item attributes with target attention, tokenizes context and target attributes, and applies Zenith interaction blocks.
\item \textbf{TokenMixer}~\cite{tokenmixer} combines target-aware sequence summaries with global and group tokens, then applies reversible token mixing and per-token feed-forward networks.
\item \textbf{UniMixer}~\cite{unimixer} flattens histories with user and context fields, then chunk-tokenizes them for UniMixing-Lite blocks.
\item \textbf{HeMix}~\cite{hemix} combines fixed and feature-conditioned queries with low-rank token mixing over recent and earlier histories.
\item \textbf{SSR}~\cite{ssr} uses target-attention sequence pooling and multi-view competitive sparse filters with dense fusion.
\end{itemize}

\subsubsection{\textbf{Layer-wise Unified Interaction}}
This paradigm integrates sequential and non-sequential representations throughout every network layer.
\begin{itemize}[leftmargin=*]
\item \textbf{OneTrans}~\cite{onetrans} injects sequence and partitioned non-sequential tokens into every block, using mixed attention, a shared sequence FFN, and token-specific FFNs.
\item \textbf{HyFormer}~\cite{hyformer} encodes the sequence per layer, decodes global queries, and enriches them with non-sequential tokens.
\item \textbf{MixFormer}~\cite{mixformer_rec} mixes non-sequential queries, uses cross-attention to retrieve information from the updated sequence, and fuses the resulting representations.
\item \textbf{INFNet}~\cite{infnet} organizes categorical, sequential, and task tokens around a hub for layer-wise aggregation and broadcasting.
\item \textbf{EST}~\cite{est} uses cross-attention to sequence representations and feature-specific feed-forward layers.
\item \textbf{TokenFormer}~\cite{tokenformer} represents fields, histories, and targets as tokens with full interaction and sliding-window Flex Attention.
\item \textbf{LONGER}~\cite{longer} encodes local groups with inner Transformers and joins compressed long-term history with recent queries.
\item \textbf{UltraHSTU}~\cite{ultrahstu} extends HSTU-style sequential transduction~\cite{zhai2024actions} with hybrid attention that combines sparse and full attention in each interaction layer.
\end{itemize}

\section{Benchmarking Ranking Models}
\subsection{Reproducibility Analysis}
Ranking models receive substantial attention from both academia and industry. Many advances are developed with proprietary data, implementations, and training infrastructure. Differences in access to these resources can make it difficult to determine whether gains arise from model architecture, data processing, training pipelines, or engineering optimization. Moreover, dataset and reproduction choices may alter model rankings even in controlled experiments. We therefore define six requirements for reproducible and fair ranking model research.

\begin{itemize}[leftmargin=*]
\item \textbf{Unified data pipeline.} Studies use consistent raw-log filtering, feature extraction, categorical encoding, sequence construction, target alignment, chronological splitting, truncation, and padding. This creates matched samples and prevents future leakage from confounding comparisons.
\item \textbf{Unified task and metric definitions.} Each feedback task uses consistent labels, sample eligibility rules, loss functions, metric implementations, and aggregation procedures. This ensures measured differences stem from architecture, not task formulation.
\item \textbf{Unified model source code.} Baselines use complete, verifiable source code and consistent interfaces. This avoids confusing implementation differences with architecture.
\item \textbf{Unified hyperparameters.} Shared capacity parameters such as network depth, token count, token dimension, embedding dimension, and sequence length should be aligned as closely as possible across baselines. This prevents capacity mismatches from driving apparent model gains, while architecture-specific parameters vary only for defining operations.
\item \textbf{Unified training pipeline.} Baselines share data loading, training, validation, and testing. The pipeline also controls optimizers, learning rates, effective batch size, accumulation, initialization, epochs, and task weighting so that training choices do not masquerade as architectural improvements.
\item \textbf{Unified efficiency optimization.} Studies align and disclose distributed training, precision, attention kernels, operator compilation, checkpointing, hardware, and software configurations. This makes performance and resource use comparable under the same systems setting.
\end{itemize}

\suppressfloats[t]
\begin{table}[t]
\Huge
\renewcommand\arraystretch{1.2}
\caption{Dataset statistics.}
\resizebox{\linewidth}{!}{
\begin{tabular}{c|ccccccc}
\hline
\textbf{Dataset}  & \textbf{\#Instances} & \textbf{\#Users} & \textbf{\#Items} & \textbf{\#Fields} & \textbf{\#Tasks} & \textbf{\#AvgL} & \textbf{\#MaxL} \\ \hline
\textbf{QK-Video} & 493,306,303          & 4,996,176        & 3,752,235        & 10               & 4               & 99                    & 6,013                 \\
\textbf{KuaiRand} & 323,464,444          & 27,285           & 32,038,725       & 40               & 6               & 11,855                & 228,030               \\
\textbf{TAAC-25}  & 757,207,146          & 7,706,778        & 15,707,425       & 30               & 2               & 98                    & 100                   \\
\textbf{Taobao}   & 23,601,301           & 470,570          & 831,643          & 23               & 4               & 50                    & 3,756                 \\
\textbf{MerRec}   & 172,304,959          & 1,697,072        & 42,577,610       & 20               & 5               & 102                   & 26,576                \\ \hline
\end{tabular}}
\label{tab:datasets}
\vspace{-0.5em}
\end{table}

Table~\ref{reproducibility} summarizes how existing studies cover the six requirements. The symbols \cmark, \pmark, and \xmark{} denote fully satisfied, unknown or partially satisfied, and not satisfied, respectively. Coverage varies across dimensions: task and metric definitions are often documented, whereas complete preprocessing pipelines, official implementations, capacity controls, training specifications, and efficiency settings are not always jointly available. Consequently, no existing study covers all six dimensions under a single public protocol. \unirank{} consolidates these requirements into a unified public benchmark for reproducible and fair comparison.

\subsection{Evaluation Protocol}
\subsubsection{\textbf{Datasets}}

\unirank{} evaluates five large-scale public datasets spanning short-video recommendation, advertising, e-commerce, and C2C marketplaces. Table~\ref{tab:datasets} summarizes their scale, feature fields, feedback tasks, and behavior lengths. The main benchmark fixes sequence length at 100 for controlled comparison across all 15 models. Dataset provenance, collection periods, raw interaction types, and task definitions are detailed in Appendix~\ref{app:dataset-descriptions}.

\subsubsection{\textbf{Data Preprocessing}}
\unirank{} maps heterogeneous platform logs to a shared pointwise schema of \textit{user}, \textit{context}, \textit{sequence}, and \textit{action} features. Each retained interaction becomes a target sample, with its preceding item--action pairs forming the behavior sequence. The pipeline joins user and item side information, encodes categorical fields, maps infrequent values to OOV, and writes matched \texttt{data}, \texttt{user\_info}, and \texttt{item\_info} Parquet blocks for efficient loading. Dataset-specific construction and preprocessing rules are detailed in Appendix~\ref{app:dataset-preprocessing}.

\subsubsection{\textbf{Data Splitting}}
\begin{figure}[t]
  \centering
  \includegraphics[width=0.8\linewidth]{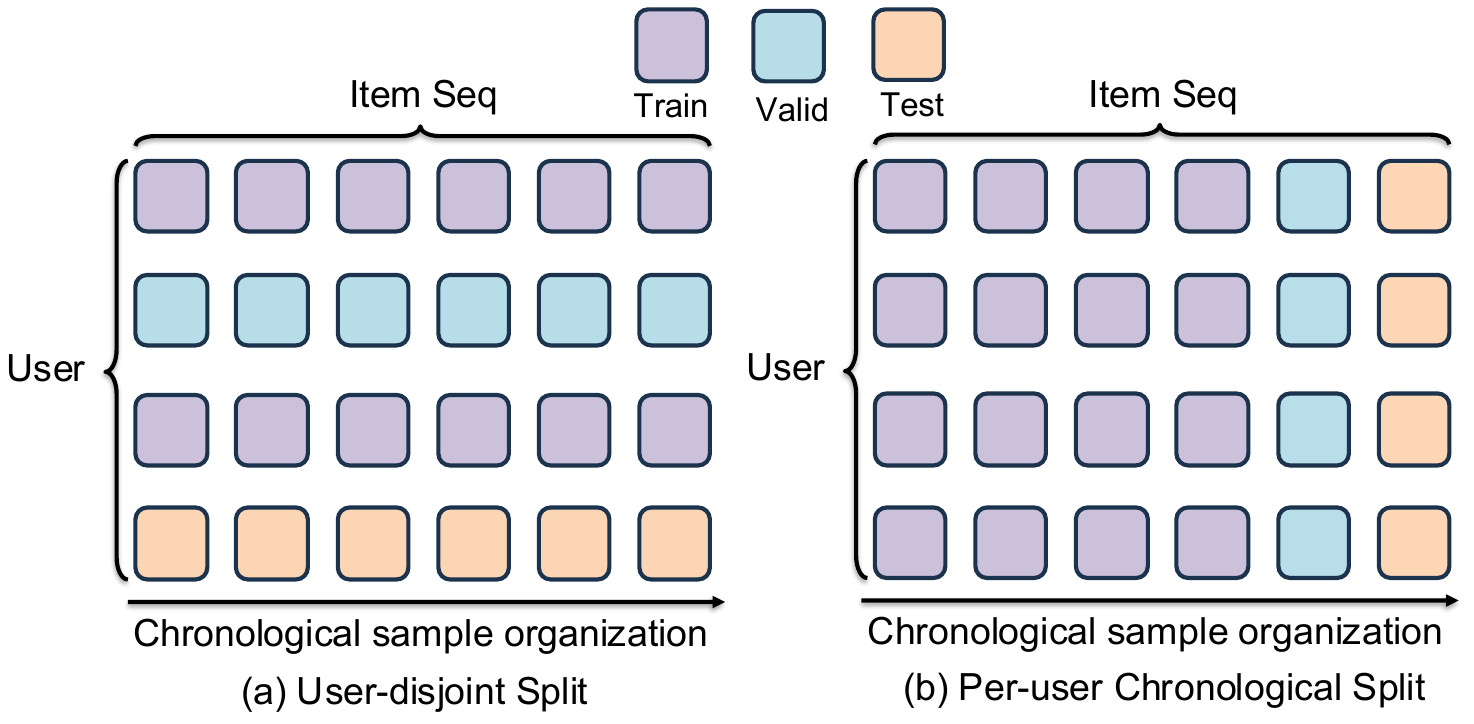}
  \caption{Comparison of randomized user-disjoint and per-user chronological sample splitting.}
  \label{fig:test-pipeline}
  \vspace{-0.5em}
\end{figure}

Figure~\ref{fig:test-pipeline} compares a randomized user-disjoint split with \unirank{}'s chronological split. The user-disjoint protocol assigns each user and their entire interaction history exclusively to one data partition. In fact, this protocol is better suited to cold-start evaluation, as it primarily measures generalization to unseen users rather than future interactions from returning users. In addition, the partitions may cover overlapping calendar periods, allowing training interactions to occur after validation or test interactions from other users. Such temporal inversion exposes the model to future popularity patterns and platform trends, making offline evaluation less representative of deployment.

\unirank{} instead orders all interactions chronologically and partitions them into training, validation, and testing sets with an approximate 8:1:1 ratio. Within each user trajectory, each target interaction is constructed using only preceding behavior. The validation period supports model configuration tuning, while the test period evaluates performance on the latest interactions. This protocol emulates past-to-future prediction by training on users' earlier interactions and evaluating their subsequent interactions, while also assessing robustness to shifts in user interests over time.

\begin{figure}[t]
  \centering
  \includegraphics[width=0.8\linewidth]{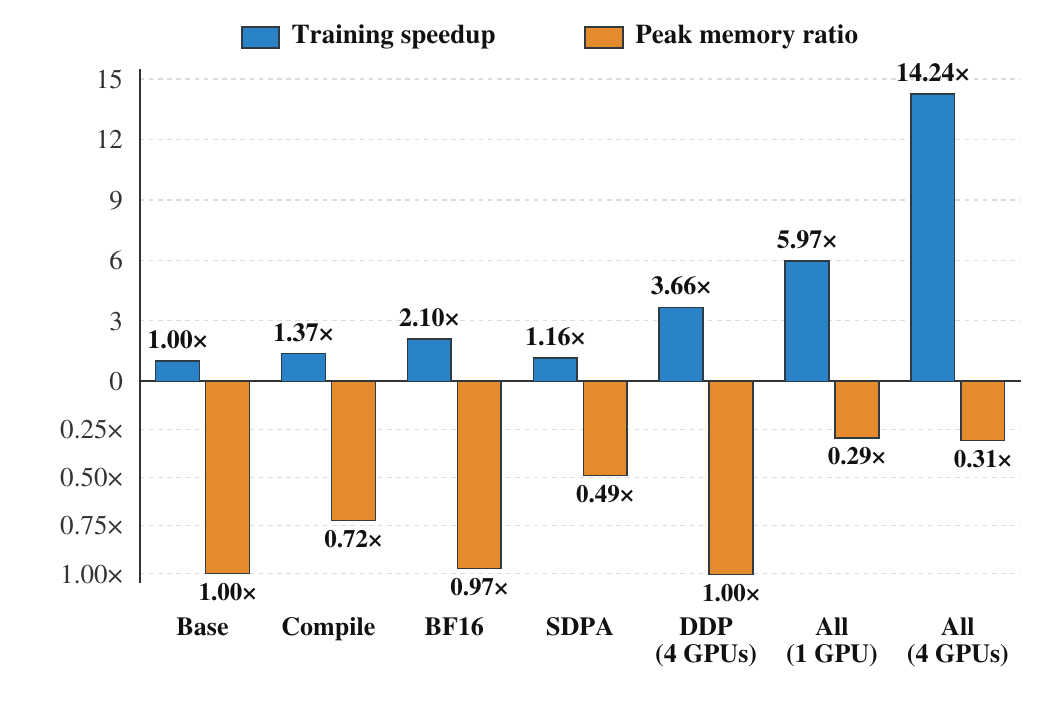}
  \caption{Training speedup and per-GPU peak memory.}
  \label{fig:training-speedup}
  \vspace{-0.5em}
\end{figure}

\subsubsection{\textbf{Evaluation Metrics}}
The benchmark reports two widely used metrics, binary Logloss and AUC, for every task~\cite{openbenchmark,FuxiCTR,shen2017deepctr}. Lower Logloss indicates better probabilistic calibration, whereas higher AUC indicates better global discrimination.

\begin{table*}[t]
\Huge
\renewcommand\arraystretch{1.4}
\caption{Comprehensive Benchmarking Results (Sequence Length = 100)}
\label{benchmark}
\resizebox{\textwidth}{!}{
\begin{tabular}{c|c|c|ccccccc|cccccccc}
\hline
\textbf{Dataset} & \textbf{Task} & \textbf{Metric} & \textbf{HiFormer} & \textbf{RankMixer} & \textbf{Zenith} & \textbf{TokenMixer} & \textbf{UniMixer} & \textbf{HeMix} & \textbf{SSR} & \textbf{OneTrans} & \textbf{LONGER} & \textbf{HyFormer} & \textbf{MixFormer} & \textbf{INFNet} & \textbf{EST} & \textbf{TokenFormer} & \textbf{UltraHSTU} \\ \hline

\multirow{8}{*}{\textbf{QK-Video}} & \multirow{2}{*}{\textbf{Click}} & \textbf{AUC} & 0.9323 & 0.9318 & 0.9334 & 0.9336 & 0.9327 & \textbf{0.9360 (2)} & 0.9314 & \textbf{0.9357 (3)} & 0.9356 & 0.9330 & 0.9333 & 0.9332 & 0.9344 & 0.9350 & \textbf{0.9363 (1)} \\
 &  & \textbf{Logloss} & 0.2539 & 0.2547 & 0.2521 & 0.2521 & 0.2533 & \textbf{0.2478 (2)} & 0.2554 & \textbf{0.2484 (3)} & 0.2486 & 0.2529 & 0.2525 & 0.2527 & 0.2505 & 0.2508 & \textbf{0.2474 (1)} \\
 & \multirow{2}{*}{\textbf{Follow}} & \textbf{AUC} & 0.9185 & 0.9108 & \textbf{0.9222 (2)} & \textbf{0.9219 (3)} & 0.9023 & \textbf{0.9227 (1)} & 0.9081 & 0.9183 & 0.9166 & 0.9110 & 0.9173 & 0.9190 & 0.9208 & 0.9163 & 0.9217 \\
 &  & \textbf{Logloss} & \textbf{0.0050 (2)} & 0.0052 & \textbf{0.0050 (2)} & \textbf{0.0051 (3)} & 0.0052 & \textbf{0.0049 (1)} & \textbf{0.0051 (3)} & \textbf{0.0051 (3)} & \textbf{0.0051 (3)} & 0.0052 & \textbf{0.0051 (3)} & \textbf{0.0051 (3)} & \textbf{0.0050 (2)} & \textbf{0.0051 (3)} & \textbf{0.0049 (1)} \\
 & \multirow{2}{*}{\textbf{Like}} & \textbf{AUC} & 0.9465 & 0.9444 & 0.9480 & 0.9475 & 0.9428 & \textbf{0.9496 (2)} & 0.9428 & 0.9483 & 0.9489 & 0.9464 & 0.9470 & 0.9482 & \textbf{0.9490 (3)} & 0.9483 & \textbf{0.9498 (1)} \\
 &  & \textbf{Logloss} & 0.0372 & 0.0375 & 0.0373 & 0.0374 & 0.0373 & \textbf{0.0361 (1)} & 0.0381 & \textbf{0.0364 (3)} & \textbf{0.0362 (2)} & 0.0370 & 0.0369 & 0.0370 & 0.0366 & \textbf{0.0364 (3)} & \textbf{0.0361 (1)} \\
 & \multirow{2}{*}{\textbf{Share}} & \textbf{AUC} & 0.9334 & 0.9290 & \textbf{0.9358 (2)} & 0.9352 & 0.9239 & \textbf{0.9368 (1)} & 0.9292 & 0.9339 & 0.9330 & 0.9305 & 0.9320 & 0.9347 & 0.9345 & 0.9327 & \textbf{0.9357 (3)} \\
 &  & \textbf{Logloss} & \textbf{0.0053 (2)} & \textbf{0.0054 (3)} & \textbf{0.0053 (2)} & \textbf{0.0053 (2)} & 0.0055 & \textbf{0.0052 (1)} & 0.0055 & \textbf{0.0053 (2)} & \textbf{0.0053 (2)} & \textbf{0.0054 (3)} & 0.0055 & \textbf{0.0053 (2)} & \textbf{0.0053 (2)} & \textbf{0.0053 (2)} & \textbf{0.0053 (2)} \\ \hline

\multirow{12}{*}{\textbf{KuaiRand}} & \multirow{2}{*}{\textbf{Click}} & \textbf{AUC} & 0.7782 & 0.7810 & 0.7816 & 0.7778 & \textbf{0.7935 (3)} & 0.7858 & 0.7872 & 0.7818 & \textbf{0.7962 (2)} & 0.7817 & 0.7795 & 0.7669 & 0.7719 & \textbf{0.7990 (1)} & 0.7744 \\
 &  & \textbf{Logloss} & 0.5400 & 0.5408 & 0.5389 & 0.5440 & \textbf{0.5253 (3)} & 0.5351 & 0.5330 & 0.5531 & \textbf{0.5226 (2)} & 0.5412 & 0.5453 & 0.5677 & 0.5747 & \textbf{0.5222 (1)} & 0.5729 \\
 & \multirow{2}{*}{\textbf{Follow}} & \textbf{AUC} & \textbf{0.8872 (2)} & 0.8739 & 0.8844 & 0.8865 & 0.8698 & \textbf{0.8869 (3)} & 0.8782 & 0.8715 & 0.8789 & 0.8800 & 0.8807 & 0.8728 & \textbf{0.8921 (1)} & 0.8762 & 0.8746 \\
 &  & \textbf{Logloss} & \textbf{0.0063 (3)} & 0.0064 & 0.0065 & \textbf{0.0062 (2)} & 0.0064 & \textbf{0.0061 (1)} & 0.0064 & 0.0064 & \textbf{0.0063 (3)} & 0.0066 & 0.0066 & 0.0064 & \textbf{0.0062 (2)} & \textbf{0.0062 (2)} & \textbf{0.0062 (2)} \\
 & \multirow{2}{*}{\textbf{Like}} & \textbf{AUC} & \textbf{0.9294 (2)} & 0.9241 & \textbf{0.9294 (2)} & 0.9270 & 0.9260 & 0.9266 & 0.9274 & 0.9215 & \textbf{0.9281 (3)} & 0.9215 & 0.9233 & 0.9207 & \textbf{0.9301 (1)} & 0.9193 & 0.9252 \\
 &  & \textbf{Logloss} & \textbf{0.0535 (1)} & 0.0553 & \textbf{0.0537 (2)} & 0.0550 & 0.0543 & 0.0545 & \textbf{0.0537 (2)} & 0.0554 & \textbf{0.0538 (3)} & 0.0559 & 0.0551 & 0.0575 & 0.0545 & 0.0578 & \textbf{0.0535 (1)} \\
 & \multirow{2}{*}{\textbf{Comment}} & \textbf{AUC} & \textbf{0.8993 (2)} & 0.8845 & \textbf{0.8994 (1)} & \textbf{0.8993 (2)} & 0.8870 & \textbf{0.8974 (3)} & 0.8886 & 0.8903 & 0.8928 & 0.8853 & 0.8894 & 0.8846 & \textbf{0.8993 (2)} & 0.8753 & 0.8949 \\
 &  & \textbf{Logloss} & \textbf{0.0147 (1)} & 0.0152 & \textbf{0.0147 (1)} & \textbf{0.0147 (1)} & \textbf{0.0150 (3)} & \textbf{0.0148 (2)} & \textbf{0.0150 (3)} & \textbf{0.0150 (3)} & \textbf{0.0150 (3)} & 0.0151 & \textbf{0.0150 (3)} & 0.0155 & \textbf{0.0148 (2)} & 0.0155 & \textbf{0.0148 (2)} \\
 & \multirow{2}{*}{\textbf{Forward}} & \textbf{AUC} & 0.8710 & 0.8499 & 0.8704 & \textbf{0.8744 (2)} & 0.8541 & \textbf{0.8738 (3)} & 0.8530 & 0.8622 & 0.8573 & 0.8691 & 0.8702 & 0.8592 & \textbf{0.8788 (1)} & 0.8537 & 0.8683 \\
 &  & \textbf{Logloss} & \textbf{0.0057 (2)} & \textbf{0.0058 (3)} & \textbf{0.0056 (1)} & \textbf{0.0056 (1)} & \textbf{0.0058 (3)} & \textbf{0.0057 (2)} & \textbf{0.0058 (3)} & 0.0061 & 0.0059 & \textbf{0.0058 (3)} & \textbf{0.0057 (2)} & 0.0060 & \textbf{0.0057 (2)} & \textbf{0.0058 (3)} & \textbf{0.0056 (1)} \\
 & \multirow{2}{*}{\textbf{Long View}} & \textbf{AUC} & 0.7961 & 0.8002 & 0.7981 & 0.7955 & \textbf{0.8111 (3)} & 0.7998 & 0.8056 & 0.8016 & \textbf{0.8127 (2)} & 0.7984 & 0.7982 & 0.7758 & 0.7935 & \textbf{0.8157 (1)} & 0.7985 \\
 &  & \textbf{Logloss} & 0.4569 & 0.4539 & 0.4582 & 0.4600 & \textbf{0.4420 (3)} & 0.4589 & 0.4475 & 0.4597 & \textbf{0.4405 (2)} & 0.4552 & 0.4561 & 0.5048 & 0.4717 & \textbf{0.4387 (1)} & 0.4654 \\ \hline

\multirow{4}{*}{\textbf{TAAC-25}} & \multirow{2}{*}{\textbf{Click}} & \textbf{AUC} & 0.8310 & 0.8069 & 0.8236 & 0.8306 & 0.8089 & 0.8249 & 0.8113 & 0.8328 & 0.8335 & 0.8267 & 0.8132 & 0.8394 & \textbf{0.8475 (2)} & \textbf{0.8532 (1)} & \textbf{0.8466 (3)} \\
 &  & \textbf{Logloss} & \textbf{0.0355 (2)} & 0.0412 & 0.0583 & 0.0393 & 0.0432 & 0.0645 & 0.0619 & 0.0439 & 0.0432 & 0.0527 & 0.0605 & 0.0501 & \textbf{0.0335 (1)} & \textbf{0.0365 (3)} & 0.0395 \\
 & \multirow{2}{*}{\textbf{Conversion}} & \textbf{AUC} & 0.8730 & 0.8683 & 0.8627 & 0.8666 & 0.8626 & 0.8712 & 0.8716 & 0.8806 & 0.8863 & 0.8758 & 0.8754 & \textbf{0.8866 (3)} & 0.8622 & \textbf{0.8871 (2)} & \textbf{0.8951 (1)} \\
 &  & \textbf{Logloss} & \textbf{0.0177 (3)} & 0.0216 & 0.0198 & 0.0191 & 0.0237 & 0.0241 & 0.0233 & 0.0232 & 0.0223 & 0.0185 & 0.0185 & \textbf{0.0173 (2)} & \textbf{0.0157 (1)} & 0.0182 & 0.0189 \\ \hline

\multirow{8}{*}{\textbf{Taobao}} & \multirow{2}{*}{\textbf{Click}} & \textbf{AUC} & \textbf{0.6446 (3)} & 0.6039 & 0.6217 & 0.6419 & 0.5999 & \textbf{0.6523 (1)} & 0.6006 & 0.6295 & 0.6377 & 0.6107 & 0.6444 & 0.6322 & \textbf{0.6478 (2)} & 0.6342 & 0.6366 \\
 &  & \textbf{Logloss} & 0.1862 & 0.1887 & 0.1874 & \textbf{0.1858 (3)} & 0.1891 & \textbf{0.1849 (1)} & 0.1890 & 0.1886 & 0.1864 & 0.1890 & \textbf{0.1857 (2)} & 0.1887 & 0.1874 & 0.1867 & 0.1869 \\
 & \multirow{2}{*}{\textbf{Cart}} & \textbf{AUC} & 0.8167 & 0.6754 & 0.7829 & 0.8253 & 0.7602 & 0.8284 & 0.7534 & 0.7458 & 0.8024 & 0.7484 & 0.8286 & \textbf{0.8366 (2)} & \textbf{0.8404 (1)} & 0.8217 & \textbf{0.8333 (3)} \\
 &  & \textbf{Logloss} & 0.0183 & 0.0207 & 0.0192 & 0.0180 & 0.0192 & \textbf{0.0178 (3)} & 0.0201 & 0.0199 & 0.0188 & 0.0198 & 0.0182 & 0.0188 & \textbf{0.0177 (2)} & \textbf{0.0178 (3)} & \textbf{0.0176 (1)} \\
 & \multirow{2}{*}{\textbf{Favor}} & \textbf{AUC} & 0.8471 & 0.7211 & 0.7993 & 0.8534 & 0.7957 & 0.8535 & 0.7816 & 0.7833 & 0.8224 & 0.7806 & \textbf{0.8558 (3)} & \textbf{0.8597 (2)} & \textbf{0.8696 (1)} & 0.8507 & 0.8544 \\
 &  & \textbf{Logloss} & 0.0144 & 0.0155 & 0.0146 & \textbf{0.0135 (3)} & 0.0145 & \textbf{0.0132 (1)} & 0.0155 & 0.0148 & 0.0139 & 0.0150 & \textbf{0.0135 (3)} & 0.0137 & \textbf{0.0132 (1)} & \textbf{0.0132 (1)} & \textbf{0.0133 (2)} \\
 & \multirow{2}{*}{\textbf{Buy}} & \textbf{AUC} & 0.8424 & 0.6856 & 0.7934 & 0.8497 & 0.7676 & 0.8533 & 0.7564 & 0.7673 & 0.8248 & 0.7485 & 0.8502 & \textbf{0.8599 (2)} & \textbf{0.8684 (1)} & 0.8483 & \textbf{0.8562 (3)} \\
 &  & \textbf{Logloss} & 0.0069 & 0.0073 & 0.0068 & \textbf{0.0064 (3)} & 0.0069 & \textbf{0.0061 (1)} & 0.0072 & 0.0070 & 0.0070 & 0.0070 & \textbf{0.0064 (3)} & 0.0068 & \textbf{0.0061 (1)} & \textbf{0.0062 (2)} & \textbf{0.0061 (1)} \\ \hline

\multirow{10}{*}{\textbf{MerRec}} & \multirow{2}{*}{\textbf{Like}} & \textbf{AUC} & 0.7518 & 0.7448 & 0.7430 & 0.7533 & \textbf{0.7650 (1)} & \textbf{0.7549 (2)} & 0.7470 & 0.7490 & 0.7534 & 0.7361 & 0.7312 & 0.7296 & \textbf{0.7543 (3)} & 0.7434 & 0.7400 \\
 &  & \textbf{Logloss} & 0.3277 & 0.3319 & 0.3306 & 0.3260 & \textbf{0.3196 (1)} & \textbf{0.3256 (2)} & 0.3296 & 0.3282 & 0.3278 & 0.3341 & 0.3433 & 0.3375 & \textbf{0.3258 (3)} & 0.3297 & 0.3343 \\
 & \multirow{2}{*}{\textbf{Cart}} & \textbf{AUC} & 0.8123 & 0.7930 & 0.8019 & \textbf{0.8133 (3)} & 0.8115 & \textbf{0.8136 (2)} & 0.8005 & 0.8078 & 0.8081 & 0.7986 & 0.8012 & 0.7947 & \textbf{0.8220 (1)} & 0.8033 & 0.8012 \\
 &  & \textbf{Logloss} & 0.0530 & 0.0539 & 0.0546 & 0.0526 & \textbf{0.0523 (1)} & \textbf{0.0524 (2)} & 0.0534 & 0.0531 & \textbf{0.0525 (3)} & 0.0538 & 0.0537 & 0.0542 & \textbf{0.0523 (1)} & 0.0534 & 0.0534 \\
 & \multirow{2}{*}{\textbf{Offer}} & \textbf{AUC} & 0.7704 & 0.7288 & 0.7712 & \textbf{0.7812 (1)} & 0.7324 & \textbf{0.7784 (2)} & 0.7591 & 0.7518 & 0.7519 & 0.7400 & 0.7563 & 0.7378 & \textbf{0.7724 (3)} & 0.7390 & 0.7625 \\
 &  & \textbf{Logloss} & 0.0242 & 0.0214 & 0.0221 & \textbf{0.0210 (2)} & 0.0223 & \textbf{0.0208 (1)} & \textbf{0.0211 (3)} & 0.0221 & 0.0213 & 0.0219 & 0.0212 & 0.0230 & 0.0224 & 0.0243 & 0.0212 \\
 & \multirow{2}{*}{\textbf{Checkout}} & \textbf{AUC} & 0.8354 & 0.8379 & 0.8503 & 0.8562 & 0.8141 & \textbf{0.8682 (1)} & 0.8473 & \textbf{0.8576 (3)} & \textbf{0.8623 (2)} & 0.8514 & 0.8552 & 0.8447 & 0.8536 & 0.8305 & 0.8558 \\
 &  & \textbf{Logloss} & 0.0127 & \textbf{0.0047 (2)} & 0.0127 & 0.0063 & 0.0068 & \textbf{0.0046 (1)} & \textbf{0.0046 (1)} & 0.0068 & \textbf{0.0047 (2)} & 0.0056 & 0.0055 & 0.0090 & 0.0121 & 0.0120 & \textbf{0.0054 (3)} \\
 & \multirow{2}{*}{\textbf{Purchase}} & \textbf{AUC} & 0.8408 & 0.8407 & 0.8495 & 0.8593 & 0.8163 & \textbf{0.8708 (1)} & 0.8552 & 0.8591 & \textbf{0.8653 (2)} & 0.8475 & 0.8561 & 0.8465 & \textbf{0.8600 (3)} & 0.8335 & 0.8564 \\
 &  & \textbf{Logloss} & 0.0077 & \textbf{0.0030 (1)} & 0.0086 & 0.0040 & 0.0044 & \textbf{0.0030 (1)} & \textbf{0.0030 (1)} & 0.0044 & \textbf{0.0031 (2)} & 0.0037 & 0.0037 & 0.0056 & 0.0082 & 0.0070 & \textbf{0.0035 (3)} \\ \hline
\end{tabular}
}
\end{table*}

\begin{figure*}[t]
  \centering
  \includegraphics[width=1\linewidth]{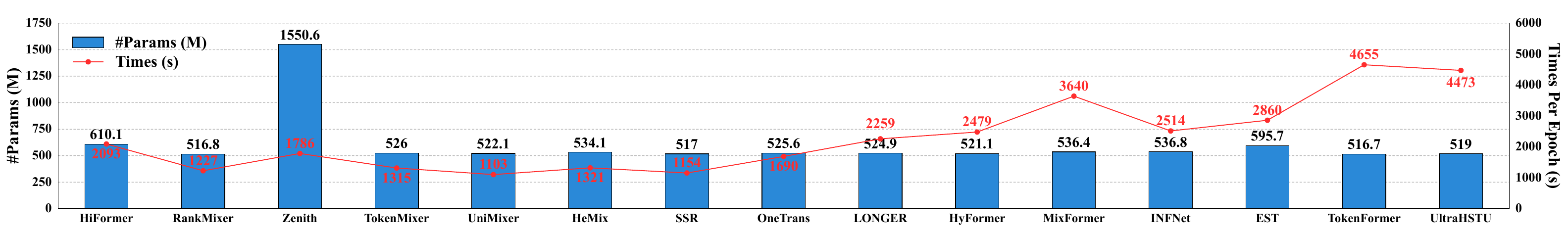}
  \caption{Efficiency comparisons on the KuaiRand dataset (Sequence Length = 100).}
  \label{fig:model-efficiency}
\end{figure*}

\subsubsection{\textbf{Toolkit Implementation of \unirank}} 
To support token-based ranking models with sequence lengths up to hundreds of thousands of items, \unirank{} provides an optimized PyTorch backend~\cite{PYTORCH}. The toolkit includes the following components:
\begin{itemize}[leftmargin=*]
\item \textbf{Distributed Data Parallel (DDP):} \unirank{} runs one CUDA process per GPU with NCCL gradient synchronization and partitions validation and testing across ranks.
\item \textbf{Operator Compilation:} \texttt{torch.compile} uses the Inductor backend for eligible dense modules while leaving sparse embedding modules outside the compiled region.
\item \textbf{Mixed Precision Training:} bf16 autocast reduces activation storage and enables Tensor Core compute, while binary cross-entropy loss is evaluated in FP32 for numerical stability.
\item \textbf{Attention Optimization:} Models using scaled dot-product attention can dispatch to fused Flash Attention when PyTorch's dtype, shape, mask, and hardware constraints are satisfied. Flex Attention supports the structured sparse attention patterns used by TokenFormer and UltraHSTU.
\item \textbf{Activation Checkpointing:} Ultra configurations use non-reentrant checkpointing for each main interaction block.
\item \textbf{Blocked Data Pipeline:} Each split uses matching \texttt{data}, \texttt{user\_info}, and \texttt{item\_info} Parquet block triplets, block-local side-information caches, and per-rank load balancing.
\item \textbf{Separate Dense and Sparse Optimization:} Dense parameters use AdamW and sparse embeddings use Adagrad with independent learning rates.
\end{itemize}

Figure~\ref{fig:training-speedup} uses OneTrans with a sequence length of 1000 and normalizes throughput and per-GPU peak memory to the one-GPU base. With all optimizations enabled, All (4 GPUs) achieves a $14.24\times$ speedup while reducing per-GPU peak memory by $69.2\%$, demonstrating the training and hardware efficiency of \unirank.

\subsubsection{\textbf{Training and Reproducibility}}
\unirank{} uses PyTorch~\cite{PYTORCH} because its imperative execution, modular interfaces, and transparent debugging simplify development and reproduction. The default batch size is 8192 per GPU. If an experiment encounters an out-of-memory error, we reduce the per-GPU batch size linearly and increase gradient accumulation by the same factor to preserve the effective batch size. Each model trains for one epoch~\cite{one-epoch} with separate sparse and dense optimizers, consistent with production practice. Adagrad optimizes sparse embeddings at a default learning rate of $0.05$, while AdamW optimizes dense parameters at $0.0001$. All experiments run on NVIDIA H20 GPUs. The main benchmark fixes the embedding dimension at 16, network depth at three layers, and token dimension at 256. All interaction blocks use pre-norm for stable optimization. These controls align capacity and optimization budgets so that performance differences primarily reflect architectural choices. A grid search tunes model-specific hyperparameters. We release the training and evaluation code, preprocessing scripts, experiment configurations, and processed datasets in the \href{https://github.com/salmon1802/UniRank}{\textcolor{blue}{GitHub repository}}, enabling end-to-end reconstruction.

\subsection{Performance Analysis}
Table~\ref{benchmark} reports results on five datasets with a sequence length of 100. Bold values and parenthesized ranks identify the top three distinct four-decimal results for each task-metric pair. Ties share a rank, and subsequent ranks remain consecutive.

\begin{itemize}[leftmargin=*]
\item \textbf{Both unified interaction paradigms are competitive:} Stacked models such as \textbf{HeMix}~\cite{hemix} and \textbf{UniMixer}~\cite{unimixer}, and layer-wise models such as \textbf{EST}~\cite{est} and \textbf{TokenFormer}~\cite{tokenformer}, lead different datasets and tasks. Neither paradigm consistently dominates under the common protocol.
\item \textbf{No universally superior model exists across data distributions:} Rankings vary substantially across QK-Video, KuaiRand, TAAC-25, Taobao, and MerRec. Platform-specific sequence patterns, feature distributions, and feedback frequencies materially affect performance, so findings may not transfer across datasets.
\item \textbf{Model performance is highly dataset-dependent:} \textbf{EST} and \textbf{HeMix}, which originate from Taobao-oriented e-commerce ranking, are competitive on Taobao and MerRec. \textbf{TokenFormer} is competitive on the advertising dataset TAAC-25. Short-video and livestreaming models, including \textbf{UltraHSTU}~\cite{ultrahstu}, \textbf{UniMixer}, and \textbf{LONGER}~\cite{longer}, are competitive on QK-Video and KuaiRand. These patterns reflect production-specific inductive biases that transfer best across similar platforms.
\item \textbf{No single model performs best across all tasks:} The strongest model changes across tasks within the same dataset. On MerRec, UniMixer, EST, and \textbf{TokenMixer}~\cite{tokenmixer} lead like, cart, and offer AUC, respectively, while HeMix leads checkout and purchase AUC. This motivates joint feedback evaluation.
\end{itemize}

\subsection{Efficiency Analysis}
Figure~\ref{fig:model-efficiency} compares total parameters and one-epoch training time under the same NVIDIA H20 GPUs, per-GPU batch size, sequence length, and precision policy. Zenith~\cite{zenith} has a substantially larger parameter footprint because it expands user and item ID embeddings and retains both IDs as independent tokens. Most of this capacity lies in sparse embedding tables, while target-aware pooling compresses the behavior sequence before interaction. Its training time therefore does not increase proportionally with parameter count. TokenFormer~\cite{tokenformer} and UltraHSTU~\cite{ultrahstu} retain sequence tokens across interaction layers and apply structured sparse attention, which increases active computation. Flex Attention can further incur kernel-launch overhead on short sequences, so its training speed is not necessarily advantageous. Dedicated kernels and smaller sparse-attention windows may improve throughput. \unirank{} prioritizes fair model comparison and tunes each model by validation AUC rather than peak system throughput. Overall, parameter count reflects mostly static representation capacity, whereas epoch time reflects activated token computation. Practical selection should therefore consider throughput, memory, storage, and sequence-length scaling together.
\subsection{A Practical Handbook}
\setlength{\floatsep}{0.3em}
\setlength{\textfloatsep}{0.5em}

\subsubsection{\textbf{Tokenization Strategies}}
\begin{itemize}[leftmargin=*,nosep]
\item \textbf{Chunk}~\cite{rankmixer} flattens the fields, divides the resulting vector into contiguous chunks, and maps each chunk to one token.
\item \textbf{Auto}~\cite{mtgr} allows each token projection to access all features.
\item \textbf{Field}~\cite{hiformer} preserves field boundaries and independently maps each field to one token.
\item \textbf{Random}~\cite{rankup} fixes a random field permutation, divides the fields into balanced groups, and maps each group to one token.
\end{itemize}
Table~\ref{tab:tokenization} shows that tokenization imposes an architectural inductive bias. Auto learns global cross-field dependencies, Field preserves semantic boundaries, and Chunk or Random promotes diverse token groups with higher informational entropy~\cite{roy2007effective}. Their relative effectiveness depends on whether interaction layers favor global access, semantic isolation, or controlled grouping.

\begin{table}[t]
\Huge
\renewcommand\arraystretch{1.2}
\caption{Model performance with different tokenization strategies}
\label{tab:tokenization}
\resizebox{\linewidth}{!}{
\begin{tabular}{c|c|c|cccc|cccc}
\hline
\multirow{2}{*}{\textbf{Dataset}}   & \multirow{2}{*}{\textbf{Task}}      & \multirow{2}{*}{\textbf{Metric}} & \multicolumn{4}{c|}{\textbf{RankMixer}}                           & \multicolumn{4}{c}{\textbf{OneTrans}}                             \\ \cline{4-11} 
                                    &                                     &                                  & \textbf{Chunk} & \textbf{Auto} & \textbf{Field} & \textbf{Random} & \textbf{Chunk} & \textbf{Auto} & \textbf{Field} & \textbf{Random} \\ \hline
\multirow{7}{*}{\textbf{KuaiRand}} & \textbf{Click} & \textbf{AUC} & 0.7810 & \textbf{0.7848} & 0.7628 & 0.7843 & 0.7574 & 0.7818 & \textbf{0.7897} & 0.7754 \\
 & \textbf{Follow} & \textbf{AUC} & \textbf{0.8739} & 0.8709 & 0.8647 & 0.8674 & 0.8557 & 0.8715 & \textbf{0.8839} & 0.8828 \\
 & \textbf{Like} & \textbf{AUC} & 0.9241 & \textbf{0.9249} & 0.9215 & 0.9248 & 0.8986 & 0.9215 & \textbf{0.9243} & 0.9242 \\
 & \textbf{Comment} & \textbf{AUC} & 0.8845 & \textbf{0.8911} & 0.8613 & 0.8885 & 0.8743 & 0.8903 & \textbf{0.8983} & 0.8909 \\
 & \textbf{Share} & \textbf{AUC} & 0.8499 & \textbf{0.8590} & 0.8501 & 0.8548 & 0.8583 & 0.8622 & \textbf{0.8762} & 0.8712 \\
 & \textbf{Long View} & \textbf{AUC} & 0.8002 & \textbf{0.8033} & 0.7798 & 0.8032 & 0.7889 & 0.8016 & \textbf{0.8049} & 0.7897 \\ \cline{2-11}
 & \textbf{avg AUC} & \textbf{AUC} & 0.8523 & \textbf{0.8557} & 0.8400 & 0.8538 & 0.8389 & 0.8548 & \textbf{0.8629} & 0.8557 \\ \hline
\multirow{6}{*}{\textbf{MerRec}} & \textbf{Like} & \textbf{AUC} & \textbf{0.7448} & 0.7397 & 0.7429 & 0.7409 & 0.7451 & 0.7490 & 0.7441 & \textbf{0.7493} \\
 & \textbf{Cart} & \textbf{AUC} & 0.7930 & 0.7879 & \textbf{0.8031} & 0.7924 & 0.8084 & 0.8078 & 0.8116 & \textbf{0.8139} \\
 & \textbf{Offer} & \textbf{AUC} & 0.7288 & 0.7271 & \textbf{0.7473} & 0.7374 & 0.7610 & 0.7518 & 0.7549 & \textbf{0.7728} \\
 & \textbf{Checkout} & \textbf{AUC} & \textbf{0.8379} & 0.8359 & 0.8378 & 0.8338 & 0.8607 & 0.8576 & 0.8560 & \textbf{0.8713} \\
 & \textbf{Purchase} & \textbf{AUC} & 0.8407 & 0.8369 & \textbf{0.8427} & 0.8376 & 0.8624 & 0.8591 & 0.8523 & \textbf{0.8748} \\ \cline{2-11}
 & \textbf{avg AUC} & \textbf{AUC} & 0.7891 & 0.7855 & \textbf{0.7948} & 0.7884 & 0.8075 & 0.8051 & 0.8038 & \textbf{0.8164} \\ \hline
\end{tabular}
}
\vspace{0.3em}
\end{table}

\begin{table}[t]
\Huge
\renewcommand\arraystretch{1.2}
\caption{Model performance with different attention activations}
\label{tab:activation}
\resizebox{\linewidth}{!}{
\begin{tabular}{c|c|c|cccccccc}
\hline
\multirow{2}{*}{\textbf{Dataset}}   & \multirow{2}{*}{\textbf{Task}}      & \multirow{2}{*}{\textbf{Metric}} & \multicolumn{8}{c}{\textbf{OneTrans}}                                                                                                   \\ \cline{4-11} 
                                    &                                     &                                  & \textbf{Base} & \textbf{None} & \textbf{ReLU} & \textbf{SoftPlus} & \textbf{SiLU} & \textbf{GeLU} & \textbf{Sigmoid} & \textbf{Mish} \\ \hline
\multirow{7}{*}{\textbf{KuaiRand}} & \textbf{Click}     & \textbf{AUC} & 0.7818 & 0.7951 & 0.7861 & 0.7859 & \textbf{0.7978} & 0.7968 & 0.7920 & 0.7961 \\
                                    & \textbf{Follow}    & \textbf{AUC} & 0.8715 & 0.8626 & 0.8712 & 0.8524 & 0.8722 & \textbf{0.8753} & 0.8731 & 0.8684 \\
                                    & \textbf{Like}      & \textbf{AUC} & 0.9215 & 0.9214 & 0.9242 & 0.9124 & 0.9231 & 0.9239 & \textbf{0.9246} & 0.9227 \\
                                    & \textbf{Comment}   & \textbf{AUC} & \textbf{0.8903} & 0.8861 & 0.8902 & 0.8662 & 0.8878 & 0.8896 & 0.8867 & 0.8894 \\
                                    & \textbf{Share}     & \textbf{AUC} & \textbf{0.8622} & 0.8519 & 0.8567 & 0.8536 & 0.8601 & 0.8603 & 0.8593 & 0.8569 \\
                                    & \textbf{Long View} & \textbf{AUC} & 0.8016 & 0.8115 & 0.8038 & 0.8001 & \textbf{0.8143} & 0.8139 & 0.8111 & 0.8127 \\ \cline{2-11}
                                    & \textbf{avg AUC}   & \textbf{AUC} & 0.8548 & 0.8548 & 0.8554 & 0.8451 & 0.8592 & \textbf{0.8599} & 0.8578 & 0.8577 \\ \hline
\multirow{6}{*}{\textbf{MerRec}}   & \textbf{Like}      & \textbf{AUC} & 0.7490 & 0.7507 & 0.7393 & 0.7510 & 0.7516 & 0.7535 & \textbf{0.7582} & 0.7473 \\
                                    & \textbf{Cart}      & \textbf{AUC} & 0.8078 & 0.8093 & 0.8086 & 0.8087 & \textbf{0.8134} & 0.8125 & 0.8087 & 0.8065 \\
                                    & \textbf{Offer}     & \textbf{AUC} & 0.7518 & 0.7617 & 0.7504 & 0.7438 & 0.7616 & \textbf{0.7625} & 0.7427 & 0.7578 \\
                                    & \textbf{Checkout}  & \textbf{AUC} & 0.8576 & 0.8586 & 0.8575 & 0.8587 & \textbf{0.8593} & 0.8588 & 0.8589 & 0.8592 \\
                                    & \textbf{Buy}       & \textbf{AUC} & 0.8591 & 0.8557 & 0.8586 & \textbf{0.8620} & 0.8566 & 0.8573 & 0.8618 & 0.8605 \\ \cline{2-11}
                                    & \textbf{avg AUC}   & \textbf{AUC} & 0.8051 & 0.8072 & 0.8029 & 0.8048 & 0.8085 & \textbf{0.8089} & 0.8061 & 0.8063 \\ \hline
\end{tabular}
}
\vspace{0.3em}
\end{table}

\subsubsection{\textbf{Attention Activations.}}
We compare the Softmax baseline~\cite{Attentionisallyouneed} with unnormalized logits~\cite{DIN}, ReLU~\cite{activations1}, SoftPlus~\cite{activations2}, SiLU~\cite{activations3}, GeLU~\cite{hendrycks2016gelu}, Sigmoid~\cite{activations2}, and Mish~\cite{misra2019mish}. Table~\ref{tab:activation} shows that smooth signed activations are the most robust. GeLU leads average AUC on both datasets and SiLU performs similarly, suggesting that signed responses outperform competitive or non-negative aggregation. SoftPlus degrades KuaiRand because weak positive weights accumulate noise. Task optima still vary: Sigmoid leads Like, whereas unnormalized logits benefit Click and Long View. Overall, GeLU and SiLU are strong performance-oriented defaults, while Softmax remains attractive when fused-kernel efficiency is important~\cite{dao2022flashattention}.

\subsubsection{\textbf{Architectural Enhancements.}}
\begin{itemize}[leftmargin=*,nosep]
\item \textbf{SpecT}~\cite{xiao2024streamingllm} adds learnable boundary tokens as attention sinks.
\item \textbf{AttGate}~\cite{qiu2025gated} applies a learned sigmoid gate after attention.
\item \textbf{RoPE}~\cite{su2024roformer} rotates sequence queries and keys to encode position.
\item \textbf{QKNorm}~\cite{henry2020query} normalizes attention queries and keys.
\item \textbf{GNorm}~\cite{wu2018groupnorm} uses heterogeneous LayerNorm: sequence tokens share one, while non-sequential tokens use individual ones.
\item \textbf{AttRes}~\cite{he2021realformer} adds residual attention logits across adjacent layers.
\item \textbf{Tau} learns a separate attention scale for each head.
\end{itemize}
Table~\ref{tab:tricks} shows that these enhancements have substantially different cross-dataset robustness. AttGate is the only component that improves every task on both datasets. It therefore provides the safest general-purpose enhancement by suppressing target-irrelevant attention updates. QKNorm and GNorm perform best on KuaiRand, but GNorm degrades MerRec, suggesting that heterogeneous normalization may remove useful magnitude differences between token groups. RoPE provides larger gains on MerRec, indicating that chronological order has dataset-dependent importance. SpecT mainly benefits KuaiRand, whereas Tau reduces its average AUC. These results show that most enhancements align with dataset-specific characteristics rather than providing universal gains.

\begin{table}[t]
\Huge
\renewcommand\arraystretch{1.2}
\caption{Model performance with architectural enhancements}
\label{tab:tricks}
\resizebox{\linewidth}{!}{
\begin{tabular}{c|c|c|cccccccc}
\hline
\multirow{2}{*}{\textbf{Dataset}}   & \multirow{2}{*}{\textbf{Task}}      & \multirow{2}{*}{\textbf{Metric}} & \multicolumn{8}{c}{\textbf{OneTrans}}                                                                                                 \\ \cline{4-11} 
                                    &                                     &                                  & \textbf{Base} & \textbf{SpecT} & \textbf{AttGate} & \textbf{RoPE} & \textbf{QKNorm} & \textbf{GNorm} & \textbf{AttRes} & \textbf{Tau} \\ \hline
\multirow{7}{*}{\textbf{KuaiRand}} & \textbf{Click}     & \textbf{AUC} & 0.7818 & 0.7892 & 0.7946 & 0.7793 & 0.7963 & \textbf{0.7972} & 0.7823 & 0.7615 \\
                                    & \textbf{Follow}    & \textbf{AUC} & 0.8715 & 0.8835 & 0.8841 & 0.8747 & 0.8882 & \textbf{0.8907} & 0.8810 & 0.8753 \\
                                    & \textbf{Like}      & \textbf{AUC} & 0.9215 & 0.9258 & \textbf{0.9295} & 0.9226 & 0.9283 & 0.9292 & 0.9245 & 0.9200 \\
                                    & \textbf{Comment}   & \textbf{AUC} & 0.8903 & 0.8942 & 0.8969 & 0.8978 & 0.8975 & \textbf{0.8982} & 0.8952 & 0.8967 \\
                                    & \textbf{Share}     & \textbf{AUC} & 0.8622 & 0.8708 & 0.8646 & 0.8742 & \textbf{0.8773} & 0.8770 & 0.8691 & 0.8705 \\
                                    & \textbf{Long View} & \textbf{AUC} & 0.8016 & 0.8044 & 0.8118 & 0.8017 & \textbf{0.8120} & 0.8084 & 0.8055 & 0.7918 \\ \cline{2-11}
                                    & \textbf{avg AUC}   & \textbf{AUC} & 0.8548 & 0.8613 & 0.8636 & 0.8584 & 0.8666 & \textbf{0.8668} & 0.8596 & 0.8526 \\ \hline
\multirow{6}{*}{\textbf{MerRec}}   & \textbf{Like}      & \textbf{AUC} & 0.7490 & 0.7423 & 0.7534 & \textbf{0.7647} & 0.7445 & 0.7424 & 0.7575 & 0.7553 \\
                                    & \textbf{Cart}      & \textbf{AUC} & 0.8078 & 0.8082 & 0.8140 & 0.8077 & \textbf{0.8192} & 0.8113 & 0.8027 & 0.7909 \\
                                    & \textbf{Offer}     & \textbf{AUC} & 0.7518 & 0.7612 & 0.7591 & \textbf{0.7642} & 0.7520 & 0.7465 & 0.7567 & 0.7634 \\
                                    & \textbf{Checkout}  & \textbf{AUC} & 0.8576 & 0.8630 & \textbf{0.8727} & 0.8611 & 0.8622 & 0.8522 & 0.8617 & 0.8669 \\
                                    & \textbf{Buy}       & \textbf{AUC} & 0.8591 & 0.8644 & \textbf{0.8758} & 0.8645 & 0.8626 & 0.8565 & 0.8640 & 0.8677 \\ \cline{2-11}
                                    & \textbf{avg AUC}   & \textbf{AUC} & 0.8051 & 0.8078 & \textbf{0.8150} & 0.8125 & 0.8081 & 0.8018 & 0.8085 & 0.8088 \\ \hline
\end{tabular}
}
\vspace{0.3em}
\end{table}

\begin{table}[t]
\Huge
\renewcommand\arraystretch{1.2}
\caption{Model performance with different optimizers}
\label{tab:optimizers}
\resizebox{\linewidth}{!}{
\begin{tabular}{c|c|c|cccccc}
\hline
\multirow{2}{*}{\textbf{Dataset}}   & \multirow{2}{*}{\textbf{Task}}      & \multirow{2}{*}{\textbf{Metric}} & \multicolumn{6}{c}{\textbf{RankMixer}}                                                               \\ \cline{4-9} 
                                    &                                     &                                  & \textbf{Base} & \textbf{Muon} & \textbf{LaProp} & \textbf{Scion} & \textbf{SOAP} & \textbf{RMSProp} \\ \hline
\multirow{7}{*}{\textbf{KuaiRand}} & \textbf{Click}     & \textbf{AUC} & 0.7810 & \textbf{0.7865} & 0.7823 & 0.7778 & 0.7797 & 0.7824 \\
                                    & \textbf{Follow}    & \textbf{AUC} & 0.8739 & 0.8735 & 0.8767 & 0.8498 & \textbf{0.8788} & 0.8686 \\
                                    & \textbf{Like}      & \textbf{AUC} & 0.9241 & 0.9252 & \textbf{0.9271} & 0.9191 & 0.9262 & 0.9243 \\
                                    & \textbf{Comment}   & \textbf{AUC} & 0.8845 & 0.8849 & 0.8937 & 0.8728 & \textbf{0.8944} & 0.8826 \\
                                    & \textbf{Share}     & \textbf{AUC} & 0.8499 & 0.8578 & \textbf{0.8715} & 0.8066 & 0.8655 & 0.8543 \\
                                    & \textbf{Long View} & \textbf{AUC} & 0.8002 & \textbf{0.8024} & 0.7994 & 0.7974 & 0.7979 & 0.8003 \\ \cline{2-9}
                                    & \textbf{avg AUC}   & \textbf{AUC} & 0.8523 & 0.8551 & \textbf{0.8584} & 0.8373 & 0.8571 & 0.8521 \\ \hline
\multirow{6}{*}{\textbf{MerRec}}   & \textbf{Like}      & \textbf{AUC} & 0.7448 & 0.7592 & 0.7463 & 0.7461 & \textbf{0.7619} & 0.7480 \\
                                    & \textbf{Cart}      & \textbf{AUC} & 0.7930 & 0.8088 & 0.8023 & 0.8021 & \textbf{0.8174} & 0.7934 \\
                                    & \textbf{Offer}     & \textbf{AUC} & 0.7288 & 0.7630 & 0.7384 & 0.7336 & \textbf{0.7800} & 0.7265 \\
                                    & \textbf{Checkout}  & \textbf{AUC} & 0.8379 & 0.8311 & 0.8402 & \textbf{0.8414} & 0.8358 & 0.8369 \\
                                    & \textbf{Buy}       & \textbf{AUC} & 0.8407 & 0.8334 & 0.8465 & 0.8453 & \textbf{0.8469} & 0.8409 \\ \cline{2-9}
                                    & \textbf{avg AUC}   & \textbf{AUC} & 0.7891 & 0.7991 & 0.7948 & 0.7937 & \textbf{0.8084} & 0.7892 \\ \hline
\end{tabular}
}
\vspace{0.3em}
\end{table}

\subsubsection{\textbf{Dense Optimizers.}}
We fix Adagrad for sparse embeddings and compare AdamW~\cite{AdamW} with Muon~\cite{liu2025muon}, LaProp~\cite{liu2020laprop}, Scion~\cite{pethick2025scion}, SOAP~\cite{vyas2025soap}, and RMSProp~\cite{tieleman2012rmsprop} for dense parameters. Table~\ref{tab:optimizers} shows that no optimizer consistently dominates across datasets. LaProp performs best on KuaiRand, with the largest gains on Comment and Share. SOAP performs best on MerRec, mainly through improvements on Cart, Offer, and Purchase. This difference suggests that optimizer effectiveness depends on dataset and task geometry. Muon provides moderate and relatively stable improvements, whereas RMSProp remains close to AdamW. Scion slightly improves MerRec but substantially degrades KuaiRand, showing that constrained updates can be distribution-sensitive. Overall, SOAP, Muon, and LaProp are relatively robust choices when accuracy is prioritized.

\begin{figure}[t]
  \centering
  \includegraphics[width=0.8\linewidth]{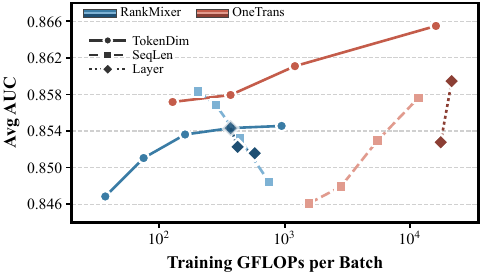}
  \caption{Scaling law study on KuaiRand.}
  \label{fig:scaling-laws-kuairand}
\end{figure}

\subsubsection{\textbf{Scaling Laws.}}
Figure~\ref{fig:scaling-laws-kuairand} presents controlled scaling experiments on KuaiRand. Under identical data, task, and training settings, we vary token dimension, sequence length, and model depth, and compare average AUC with per-batch computation. RankMixer benefits primarily from wider token representations because its target-aware pooling compresses the behavior sequence before feature interaction, limiting the gains from longer sequences or additional layers. OneTrans progressively reduces active sequence tokens across layers while retaining non-sequential tokens. Its depth scaling must therefore be coupled with the initial sequence length. Short histories leave later layers too few sequence tokens to refine, whereas longer histories preserve sufficient information after progressive reduction. These results show that effective scaling should prioritize architecture-specific dimensions according to how sequential information is preserved, compressed, and propagated, rather than uniformly increasing width, sequence length, and depth. When no clear architectural preference is evident, increasing the token dimension provides a reliable default.

\section{Related Work}
\noindent\textbf{Ranking Models.} The development of ranking models follows four broad stages: shallow feature interaction, deep feature interaction, sequence modeling, and unified modeling. Early shallow methods assign weights to individual features or capture low-order interactions through factorization-based formulations~\cite{LR,FM,FFM,FwFM,AFM}. Deep feature interaction methods subsequently embed sparse categorical features and replace manual feature crosses with neural interaction modules. Some studies learn implicit interactions through multilayer perceptrons or product transformations, while others explicitly construct high-order interactions through cross networks, self-attention, graph propagation, or polynomial functions~\cite{deepfm,pnn1,xdeepfm,dcn,dcnv2,autoint,graphfm,fignn,QNN,FCN}. Further studies reduce manual design through automatic interaction selection, feature grouping, and architecture search~\cite{autofis,deeplight,AutoPI,AutoGroup,NAS-CTR}. Despite their effectiveness, these methods primarily treat each impression as an isolated set of non-sequential fields, making it difficult to capture the temporal evolution of user interests from behavior histories.

Sequence modeling methods address this limitation by aggregating behavior histories according to the target item and tracking the temporal evolution of user interest~\cite{DIN,DIEN,mind}. Multi-task methods further share information across clicks, conversions, and other feedback signals through shared representations, expert routing, and task-specific parameters~\cite{ESMM,mmoe,PEPNET,HMoE}. However, these systems commonly combine separate sequence modules, feature interaction modules, and task towers, resulting in heterogeneous architectures in which sequence modeling and feature interaction remain loosely coupled. Recent unified ranking methods instead represent behavior sequences, feature fields, targets, and task queries through a common token interface. Some studies perform sequence modeling before feature interaction~\cite{hiformer,rankmixer,zenith,tokenmixer,unimixer,hemix,ssr}, whereas others integrate both operations within each interaction layer~\cite{onetrans,longer,hyformer,mixformer_rec,infnet,est,tokenformer,ultrahstu}. By replacing heterogeneous fine-grained modules with a common computational interface, this progression improves MFU and yields more favorable scaling curves as sequence length and model capacity increase~\cite{onerec_report,rankmixer}.

\noindent\textbf{Benchmarking.} Existing recommendation benchmarks provide complementary forms of standardization. FuxiCTR~\cite{openbenchmark,FuxiCTR} and DeepCTR~\cite{shen2017deepctr} focus on single-task CTR models evaluated on conventional tabular datasets, while RecBole~\cite{zhao2021recbole} supports a broader range of general and sequential recommendation. Scenario-Wise Rec~\cite{scenariowiserec2025} and MMLRec~\cite{yuan2024mmlrec} further extend evaluation to multi-scenario or multi-task prediction, but offer limited support for industrial-scale behavioral sequences and token-based unified architectures. Overall, these benchmarks do not jointly standardize chronological sample construction, multi-task supervision, capacity-controlled model comparison, production-oriented training protocols, and efficiency optimization. To the best of our knowledge, \unirank{} is the first benchmark specifically designed for token-based unified ranking models. It integrates large-scale public datasets, long behavior sequences, multi-task supervision, controlled experimental configurations, and optimized training pipelines within a unified framework.

\vspace{-1\baselineskip}
\section{Conclusion and Future Work}
This paper proposed \unirank, an open benchmark for ranking models with unified sequential modeling and feature interaction. \unirank{} provided standardized evaluation protocols, an optimized PyTorch toolkit, and five large-scale public datasets from diverse platforms. The evaluation covered over 700 million instances and sequences exceeding $10^5$ interactions, while handbook experiments examined design choices under limited compute. Its chronological pointwise autoregressive paradigm and acceleration components supported reproducible comparison of modern ranking architectures and enabled practical studies of scaling laws. We reported results for 15 unified ranking models and established a foundation for more open research on modern recommender systems. Future versions will extend \unirank{} to multimodal ranking models and datasets such as TaoBao-MM~\cite{taobaomm} and MicroLens~\cite{microlens}. We will also add MerRec-1B~\cite{merrec}, RecFlow~\cite{recflow}, and VK-LSVD~\cite{vklsvd}.

\bibliographystyle{ACM-Reference-Format}
\bibliography{sample-base}

\appendix
\section{Appendix}
\label{app:dataset-details}

\subsection{Dataset Descriptions}
\label{app:dataset-descriptions}

Table~\ref{tab:dataset-descriptions} inventories the source fields, ordering signals, and prediction targets used by \unirank{}. Processed examples retain global IDs, original user IDs, and available history length.

\begin{itemize}[leftmargin=*,nosep]
\item \textbf{QK-Video}~\cite{tenrec} is the short-video subset of TenRec from Tencent's recommendation feed on QQ Kandian, a news and short-video content platform. It contains ordered impressions from September 17 to December 7, 2021. Each row represents an exposed user--video pair. The four binary labels indicate whether the user clicks the video, follows its creator, likes it, or shares it. An all-zero row represents an exposure without any of these actions. The release also contains user gender and age, the number of watching behaviors on the video, and the video's content category. Its scale and strongly skewed user activity support multi-task ranking over heterogeneous histories. Because timestamps are unavailable, the source row order is the only temporal signal.

\item \textbf{KuaiRand}~\cite{kuairand} comes from Kuaishou, the short-video platform of Kuaishou Technology, and covers April 8 to May 8, 2022. It mixes ordinary recommendation traffic with random-intervention traffic, reducing policy-induced exposure bias. Each exposure includes millisecond time, request-tab context, play time, and six labels. \texttt{is\_click} denotes a click in the two-column interface or a valid play in the single-column interface. \texttt{is\_like}, \texttt{is\_follow}, \texttt{is\_comment}, and \texttt{is\_forward} respectively denote liking, following the creator, commenting, and forwarding. \texttt{long\_view} denotes completion for videos no longer than 18 seconds or at least 18 seconds of viewing for longer videos. Rich user profiles, video metadata, dense feedback, and exceptionally long histories make this dataset suitable for sequential and multi-task studies.

\item \textbf{TAAC-25}~\cite{tencentgr} uses industrial traffic from Tencent's 2025 Advertising Algorithm Competition. The product surface and feature semantics are intentionally deidentified. The timestamps in the processed data span October 15, 2024 to August 5, 2025. Raw action type 0 denotes exposure, type 1 denotes an ad click, and type 2 denotes a downstream conversion whose business meaning is anonymized. We label both types 1 and 2 as click and label type 2 as conversion, so every conversion is also a click. Four scalar user fields, four list-valued user fields, and thirteen item fields retain their anonymous numeric IDs. The resulting click--conversion funnel and production-scale traffic provide the largest instance set in the benchmark.

\item \textbf{Taobao}~\cite{taobao} comes from Alibaba Group's Taobao e-commerce and display-advertising platform. It links ad impressions, user profiles, ad features, and shopping behavior logs. Impressions span May 6--13, 2017, while the behavior log covers April 22--May 13, 2017. \texttt{is\_click} indicates whether the user clicks an exposed ad. \texttt{cart}, \texttt{fav}, and \texttt{buy} indicate whether the user subsequently adds a matched product to the cart, favorites it, or purchases it. User fields describe demographic and consumption segments, whereas ad fields identify category, campaign, advertiser, brand, and price. The linked logs therefore capture both immediate ad response and delayed shopping intent.

\item \textbf{MerRec}~\cite{merrec} comes from Mercari, Inc.'s consumer-to-consumer marketplace. The complete release contains timestamped events from May through October 2023. \unirank{} uses the largest October partition. Each sample is anchored at an item view. \texttt{Like} and \texttt{Cart} denote liking and adding that item to the cart, \texttt{Offer} denotes submitting a price offer, \texttt{Checkout} denotes starting the purchase flow, and \texttt{Purchase} denotes completing it. Product metadata includes a three-level category hierarchy, brand, condition, size, shipping party, color, and price. Unlike feed and advertising logs, its negative base is a retained view without the corresponding downstream action.
\end{itemize}

\begin{table*}[t]
\caption{Input fields and full-data label ratios normalized by the negative label.}
\label{tab:dataset-descriptions}
\footnotesize
\renewcommand{\arraystretch}{1.0}
\setlength{\tabcolsep}{3pt}
\newcommand{\datasetcell}[2]{\parbox[c][#1\baselineskip][c]{\linewidth}{\centering #2}}
\begin{tabularx}{\textwidth}{@{}>{\centering\arraybackslash}p{0.08\textwidth}>{\centering\arraybackslash}p{0.21\textwidth}>{\centering\arraybackslash}p{0.15\textwidth}>{\centering\arraybackslash}p{0.25\textwidth}>{\centering\arraybackslash}X@{}}
\toprule
\textbf{Dataset} & \textbf{User fields} & \textbf{Context fields} & \textbf{Item and sequence fields} & \textbf{Label ratio} \\
\midrule
\datasetcell{5}{\textbf{QK-Video}} & \datasetcell{5}{\texttt{user\_id}\\ \texttt{gender}\\ \texttt{age}\\ \texttt{watching\_times}} & \datasetcell{5}{Source impression order\\ No timestamp\\ No request context} & \datasetcell{5}{\texttt{item\_id}\\ \texttt{video\_category}\\ preceding items\\ four-bit action tokens} & \datasetcell{5}{Exposure: $1\times$\\ Click: $0.405\times$\\ Follow: $0.00174\times$\\ Like: $0.02098\times$\\ Share: $0.00229\times$} \\
\midrule
\datasetcell{26}{\textbf{KuaiRand}} & \datasetcell{26}{\texttt{user\_active\_degree}\\ \texttt{is\_lowactive\_period}\\ \texttt{is\_live\_streamer}\\ \texttt{is\_video\_author}\\ \texttt{follow\_user\_num\_range}\\ \texttt{fans\_user\_num\_range}\\ \texttt{friend\_user\_num\_range}\\ \texttt{register\_days\_range}\\ \texttt{onehot\_feat0}\\ \texttt{onehot\_feat1}\\ \texttt{onehot\_feat2}\\ \texttt{onehot\_feat3}\\ \texttt{onehot\_feat4}\\ \texttt{onehot\_feat5}\\ \texttt{onehot\_feat6}\\ \texttt{onehot\_feat7}\\ \texttt{onehot\_feat8}\\ \texttt{onehot\_feat9}\\ \texttt{onehot\_feat10}\\ \texttt{onehot\_feat11}\\ \texttt{onehot\_feat12}\\ \texttt{onehot\_feat13}\\ \texttt{onehot\_feat14}\\ \texttt{onehot\_feat15}\\ \texttt{onehot\_feat16}\\ \texttt{onehot\_feat17}} & \datasetcell{26}{\texttt{tab}\\ \texttt{day\_of\_week}\\ \texttt{is\_weekend}\\ \texttt{hour}\\ \texttt{time\_ms} (ordering)} & \datasetcell{26}{\texttt{video\_id}\\ \texttt{video\_type}\\ \texttt{primary\_tag}\\ \texttt{music\_type}\\ \texttt{duration\_bucket}\\ preceding items\\ six-bit actions} & \datasetcell{26}{Exposure: $1\times$\\ Click: $0.608\times$\\ Follow: $0.00106\times$\\ Like: $0.0169\times$\\ Comment: $0.00251\times$\\ Forward: $0.000871\times$\\ Long view: $0.351\times$} \\
\midrule
\datasetcell{15}{\textbf{TAAC-25}} & \datasetcell{15}{\texttt{103} (scalar)\\ \texttt{104} (scalar)\\ \texttt{105} (scalar)\\ \texttt{109} (scalar)\\ \texttt{106} (list)\\ \texttt{107} (list)\\ \texttt{108} (list)\\ \texttt{110} (list)\\ lists capped at five values} & \datasetcell{15}{\texttt{day\_of\_week}\\ \texttt{is\_weekend}\\ \texttt{hour}\\ timestamp (ordering)} & \datasetcell{15}{\texttt{100}\\ \texttt{101}\\ \texttt{102}\\ \texttt{112}\\ \texttt{114}\\ \texttt{115}\\ \texttt{116}\\ \texttt{117}\\ \texttt{118}\\ \texttt{119}\\ \texttt{120}\\ \texttt{121}\\ \texttt{122}\\ preceding items\\ exposure/click/conversion actions} & \datasetcell{15}{Exposure: $1\times$\\ Click: $0.0744\times$\\ Conversion: $0.0336\times$} \\
\midrule
\datasetcell{8}{\textbf{Taobao}} & \datasetcell{8}{\texttt{cms\_segid}\\ \texttt{cms\_group\_id}\\ \texttt{final\_gender\_code}\\ \texttt{age\_level}\\ \texttt{pvalue\_level}\\ \texttt{shopping\_level}\\ \texttt{occupation}\\ \texttt{new\_user\_class\_level}} & \datasetcell{8}{\texttt{pid}\\ \texttt{is\_weekend}\\ \texttt{hour}\\ impression time (ordering)} & \datasetcell{8}{\texttt{adgroup\_id}\\ \texttt{cate\_id}\\ \texttt{campaign\_id}\\ \texttt{customer\_id}\\ \texttt{brand}\\ \texttt{price\_bucket}\\ preceding ads\\ four-bit actions} & \datasetcell{8}{Exposure: $1\times$\\ Click: $0.0520\times$\\ Cart: $0.00386\times$\\ Favorite: $0.00284\times$\\ Buy: $0.00150\times$} \\
\midrule
\datasetcell{12}{\textbf{MerRec}} & \datasetcell{12}{\texttt{user\_id}} & \datasetcell{12}{\texttt{session\_id}\\ \texttt{day\_of\_week}\\ \texttt{is\_weekend}\\ \texttt{hour}\\ \texttt{stime} (ordering)} & \datasetcell{12}{\texttt{product\_id}\\ \texttt{price\_bucket}\\ \texttt{c0\_id}\\ \texttt{c1\_id}\\ \texttt{c2\_id}\\ \texttt{brand\_id}\\ \texttt{item\_condition\_id}\\ \texttt{size\_id}\\ \texttt{shipper\_id}\\ \texttt{color}\\ preceding view anchors\\ five-bit actions} & \datasetcell{12}{View: $1\times$\\ Like: $0.151\times$\\ Cart: $0.0128\times$\\ Offer: $0.00492\times$\\ Checkout: $0.000871\times$\\ Purchase: $0.000553\times$} \\
\bottomrule
\end{tabularx}
\vspace{0.1em}
\noindent The $1\times$ entry denotes the baseline exposure sample, while $0.151\times$, for example, indicates that the corresponding action count is 15.1\% of the baseline count.
\vspace{0em}
\end{table*}
\FloatBarrier

\twocolumn[
\begin{@twocolumnfalse}
\centering
\captionof{table}{Dataset-specific sample construction, feature processing, and chronological splitting.}
\label{tab:dataset-preprocessing}
\footnotesize
\renewcommand{\arraystretch}{1.16}
\setlength{\tabcolsep}{3.5pt}
\begin{tabular}{@{}>{\centering\arraybackslash}m{0.10\textwidth}>{\centering\arraybackslash}m{0.29\textwidth}>{\centering\arraybackslash}m{0.29\textwidth}>{\centering\arraybackslash}m{0.278\textwidth}@{}}
\toprule
\textbf{Dataset} & \textbf{Sample and label construction} & \textbf{Feature and action processing} & \textbf{Ordering and split} \\
\midrule
\textbf{QK-Video} & Retain each impression and pack its four binary labels into one history action. & Encode category, watching count, gender, and age with frequency-based OOV filtering. & Preserve per-user source order because date-based splitting is unavailable. \\
\midrule
\textbf{KuaiRand} & Treat each exposure as a sample and pack six labels into one history action. & Derive calendar context, bucket video duration, and join user and video attributes. & Sort by millisecond time and use 26 training, 3 validation, and 3 test days. \\
\midrule
\textbf{TAAC-25} & Remove all-negative users. Raw conversion events receive both conversion and click labels. & Cap list fields at five values and OOV-filter rare user and item values. & Use 226 training, 1 validation, and 68 test days with positive-label constraints. \\
\midrule
\textbf{Taobao} & Anchor on ad impressions and attribute matched cart, favorite, and buy events within 24 hours. & Join user/ad features, derive time context, bucket price, and encode four-label actions. & Sort impressions by time and use 6 training, 1 validation, and 1 test day. \\
\midrule
\textbf{MerRec} & Anchor on the first item view and attach later downstream events as labels. & Derive session/time context, bucket price, encode item fields, and pack five-label actions. & Sort view anchors by time and use 25 training, 3 validation, and 3 test days. \\
\bottomrule
\end{tabular}
\vspace{0.3em}
\end{@twocolumnfalse}
]

\subsection{Dataset Preprocessing}
\label{app:dataset-preprocessing}

All releases are converted to the same pointwise sequential format before training. By default, we remove users with fewer than ten eligible events and reserve 0 for padding and OOV. All datasets use \texttt{min\_count}=2 as the OOV threshold. When computing or memory resources are limited, this threshold can be increased to 10 to reduce vocabulary size and embedding-table memory. Global user and item IDs remain lossless, and the main benchmark fixes every feature embedding at 16 dimensions. Rather than treating raw magnitudes as continuous inputs, we discretize dataset-specific duration and price values and derive calendar context from timestamps. For label encoding, we follow ESMM~\cite{ESMM} and represent feedback labels as a multi-hot vector, allowing one sample to activate multiple tasks. Each history event pairs its item with an action token mapped from the complete multi-hot pattern. The output consists of matched \texttt{data}, \texttt{user\_info}, and \texttt{item\_info} Parquet blocks, with complete item, action, and available timestamp sequences stored in \texttt{user\_info}. Table~\ref{tab:dataset-preprocessing} summarizes these rules.

\begin{itemize}[leftmargin=*,nosep]
\item \textbf{QK-Video.} Each impression supplies four user fields, two item-related fields, and click, follow, like, and share labels. These binary labels form one multi-hot target, allowing concurrent feedback to remain positive. We map the 16 possible label combinations to history-action tokens. We keep separate global user and video vocabularies and place video category in \texttt{item\_info}. Because the release has no timestamps, an ingestion index preserves per-user source order for the chronological split. A day-based split is unavailable.

\item \textbf{KuaiRand.} We merge standard-policy and random-intervention traffic and join 26 user-profile fields with video type, first tag, music type, and duration. We convert millisecond timestamps to Asia/Shanghai date, weekday, weekend, and hour context. Video duration is discretized into seven intervals, from zero-length through longer than 60 seconds, and video metadata uses a separate categorical vocabulary. Click, follow, like, comment, forward, and long view form one multi-hot target, whose complete label combination maps to a history-action token. Chronological splitting uses 26 training days, 3 validation days, and 3 test days.

\item \textbf{TAAC-25.} Each event contains four scalar and four list-valued user fields, thirteen item fields, and an anonymous action type. Type 0 is negative for both tasks, type 1 activates only click, and type 2 activates both click and conversion. This multi-hot organization preserves conversion as a subset of click. We remove users without any positive outcome and pad or truncate each list to five values. The shortest trailing interval containing both positive labels provides 68 test days, the shortest adjacent preceding interval provides 1 validation day, and the preceding 226 days form training.

\item \textbf{Taobao.} Each ad impression anchors one example, and its \texttt{clk} value supplies the click label. We join eight user-profile fields and five ad attributes, then match cart, favorite, and purchase events from the same user, category, and brand within the following 24 hours. Click and the three downstream actions form one multi-hot target, and its complete label combination maps to a history-action token. Price is discretized into ten intervals, timestamps yield Beijing-time weekend and hour context, and chronological splitting uses 6 training days, 1 validation day, and 1 test day.

\item \textbf{MerRec.} Events are sorted by user, timestamp, and source-row order, and only the earliest \texttt{item\_view} for each user--item pair anchors a sample. Later like, cart, offer, checkout, and purchase events activate the corresponding entries in one multi-hot target but do not become separate samples or history tokens. The complete label combination maps to a history-action token. We retain session context and ten item fields, discretize price, and derive calendar fields. Chronological splitting uses 25 training days, 3 validation days, and 3 test days.
\end{itemize}
\FloatBarrier

\end{document}